\documentclass[11pt]{article}

\usepackage{textcomp}
\usepackage{pdfsync}
\usepackage{fancyhdr}
\usepackage{amssymb}
\usepackage{amsmath}
\usepackage{mathrsfs} 
\usepackage{srcltx}
\usepackage{graphicx}
\usepackage{latexsym}
\usepackage{appendix}

\def\centeron#1#2{{\setbox0=\hbox{#1}\setbox1=\hbox{#2}\ifdim
   \wd1>\wd0\kern.48\wd1\kern-.48\wd0\fi
   \copy0\kern-.48\wd0\kern-.48\wd1\copy1\ifdim\wd0>\wd1
   \kern.48\wd0\kern-.48\wd1\fi}}

\newcommand{\beq}{\begin{equation}}
\newcommand{\eeq}{\end{equation}}
\newcommand{\bea}{\begin{eqnarray}}
\newcommand{\eea}{\end{eqnarray}}
\newcommand{\ba}{\begin{array}}
\newcommand{\ea}{\end{array}}

\begin{document}


\hskip12cm
\vskip3cm

\begin{center}
 \LARGE \bf  Mass and angular momentum of black holes in 3D gravity theories with first order formalism 
\end{center}

\vskip2cm

\centerline{\Large Soonkeon Nam\footnote{nam@khu.ac.kr}\,,
~~Jong-Dae Park\footnote{jdpark@khu.ac.kr} 
}
\hskip2cm

\begin{quote}
Department of Physics and Research Institute of Basic Science, Kyung
Hee University, Kyungheedae-ro 26, Dongdaemun-gu, Seoul 02447, Seoul 130-701, Korea$^{1,2}$
\end{quote}

\hskip2cm

\vskip1cm 

\centerline{\bf Abstract} 
We apply the Wald formalism to obtain masses and angular momenta of black holes in three dimensional gravity theories using the first order formalism. Wald formalism suggests that the entropy of a black hole can be defined by an integration of a conserved charge on the bifurcation horizon,
and mass and angular momentum of a black hole as an integration of some charge variation form at spatial infinity.
The action of three dimensional gravity theories can be represented by a form including some auxiliary fields. As well-known examples we have calculated masses and angular momenta of some black holes in topologically massive gravity and new massive gravity theories using the first order formalism. 
We have also calculated mass and angular momentum of BTZ black hole and new type black hole in minimal massive gravity theory with the action represented by the first order formalism. We have also calculated the entropy and central charges of new type black hole. According to $AdS/CFT$ correspondence we suggest that the left and right moving temperatures should be equal to the Hawking temperature in the case of new type black hole in minimal massive gravity.

\thispagestyle{empty}
\renewcommand{\thefootnote}{\arabic{footnote}}
\setcounter{footnote}{0}
\newpage

\renewcommand{\theequation}{\arabic{section}.\arabic{equation}}
\setcounter{section}{0}\setcounter{equation}{0}

\section{Introduction}
For last few decades there have been paid lots of attention to three dimensional gravity theories. Studying for three dimensional gravity
theories provides an arena to explain some conceptual feature of the realistic four dimensional general relativity and some fundamental issues 
of quantum gravity. In general three dimensional spacetime of the Einstein gravity theory has no propagating degrees of freedom 
\cite{Deser:1983tn}. There only exists a black hole solution, i.e. BTZ black hole, with the negative cosmological constant \cite{Banados:1992wn,Banados:1992gq}.
A well known modification of Einstein's gravity theory in three dimensions is the topologically massive gravity theory (TMG) which consists of 
Einstein-Hilbert term, cosmological constant and the gravitational Chern-Simons term \cite{Deser:1981wh,Deser:1981wh-1,Deser:1982}, breaking parity symmetry  
with a new mass scale parameter. The linearization of this theory describes the existence of a single massive graviton mode. 
This theory also allows some black hole solutions having $AdS$ asymptotics \cite{Bouchareb:2007yx,Hotta:2008yq}. There have been many 
investigations for this TMG theory from the viewpoint of $AdS/CFT$ correspondence \cite{Skenderis:2009nt,Henningson:1998gx,Balasubramanian:1999re,deHaro:2000vlm}.

A few years ago there was a new proposition, new massive gravity (NMG) theory, which is composed of Ricci scalar with the cosmological constant and some 
specific combination of Ricci scalar square and Ricci tensor square \cite{Bergshoeff:2009hq}. The original aim of the introduction of higher curvature terms in NMG theory is to present the non-linear completion of the Pauli-Fierz  theory for massive spin 2 fields. This NMG theory has the parity symmetry and two propagating 
massive graviton modes contrary to TMG theory. There have been found some black hole solutions such as BTZ, warped AdS and new type black hole in NMG theory with a negative cosmological constant \cite{Bergshoeff:2009aq,Clement:2009gq,Oliva:2009ip,Giribet:2009qz,Ghodsi:2010gk}.
It is natural to consider the existence of the holographically dual conformal field theory (CFT) on the boundary if we could get an AdS solution in some gravity theory \cite{Maldacena:1997re-1,Maldacena:1997re,Witten:1998qj,Gubser:1998bc}. In the viewpoint of $AdS/CFT$ correspondence there was an attempt to extend this NMG theory to a theory having higher 
than square curvature terms to be consistent with the holographic $\it c$-theorem \cite{Sinha:2010ai,Gullu:2010pc,Myers:2010tj}.

In the context of $AdS/CFT$ correspondence there exists an inconsistent problem between any three dimensional gravity theories having asymptotic AdS geometry and its dual CFT on the boundary. The central charge of a dual boundary CFT becomes negative whenever the spin-2 graviton 
modes propagating on the bulk have positive energy, implying the dual CFT's non-unitarity. It is closely related to a problem that the asymptotic $AdS_3$ black hole 
solution have negative mass value whenever the bulk graviton modes have positive energy, so-called ``bulk vs. boundary clash''. 
In order to circumvent this inconsistency there were some suggestions like the Boulware-Deser ghost \cite{Boulware:1973my}, 
``Zwei Dreibein Gravity (ZDG)'' as a viable alternative to NMG \cite{Bergshoeff:2013xma,Bergshoeff:2014bia}. 
Recently the new ``minimal massive gravity(MMG)" theory has been suggested to resolve this inconsistency and to represent an alternative to TMG \cite{Bergshoeff:2014pca}. The field equation of MMG theory is distinguished from the TMG's one, including the additional symmetric curvature-squared terms while preserving 
the single bulk graviton mode state. In the action level this MMG theory can be represented by a ``Chern-Simons-like formulation'' \cite{Bergshoeff:2014bia}.  
The action of the MMG theory include the torsion term coupled with an auxiliary field $h$ which have the same odd-parity and dimension of mass with the 
spin connection $\omega$, and another parity-even `$ehh$' term considering $h$-squared term with a dimensionless parameter $\alpha$. 
For more details see some references and therein \cite{Bergshoeff:2014pca,Afshar:2014ffa,Arvanitakis:2014yja,Arvanitakis:2014xna,Merbis:2014vja}.

In spite of its difficulties there have been many studies to obtain mass and angular momentum of black holes on the curved background in three dimensional 
gravity theories. Arnowitt-Deser-Misner (ADM) have suggested a well-known method to give some conserved charges in general relativity, which describes 
an surface integral with linearized metric at infinity in asymptotically flat spacetime \cite{Arnowitt:1962hi}. Another method to obtain mass and angular momentum 
is to consider the integration of the stress-energy tensors with their counterterms on the asymptotic $AdS$ boundary surface using Brown-York formalism \cite{Balasubramanian:1999re,Brown:1992br}. Another formalism has evolved to an extended formalism, so-called Abott-Deser-Tekin (ADT) formalism, including higher curvature gravity theories \cite{Abbott:1981ff,Deser:2002rt,Deser:2002jk}. Wald has proposed a method how to calculate conserved quantities using Noether charge and symplectic potential on the covariant phase space establishing the first law of black hole thermodynamics in any covariant gravity theory 
\cite{Wald:1993nt,Jacobson:1993vj,Iyer:1994ys,Wald:1999wa}. The ADT charge is given by a surface integral of an antisymmetric ADT potential at spatial infinity, which is related to a conserved current that is described by a linearized field equation around an associated constant background contracted with a background 
Killing vector.
There was a different suggestion to obtain mass and angular momentum in asymptotically $AdS$ spacetime using $AdS/CFT$ correspondence \cite{Anninos:2008fx}. This explains that according to $AdS/CFT$ correspondence mass and angular momentum can be obtained by some combination of ${\cal E}_L$ and 
${\cal E}_R$ which express left and right energies of dual CFT. These energies can be represented 
by terms of left and right central charges and their temperatures respectively. An interesting method to find quasi-local conserved charges can be represented 
using the relation between off-shell ADT potential and linearized Noether potential \cite{Kim:2013zha,Kim:2013cor}.  There are many works to obtain 
mass and angular momentum of black holes on $AdS$ background using methods mentioned above \cite{Anninos:2008fx,Kim:2013zha,Kim:2013cor,Miskovic:2009kr,Giribet:2010ed,Hohm:2010jc,Nam:2010dd,Nam:2010ub,Deser:2003vh,Clement:2003sr,Moussa:2003fc}. 
Also there have been some studies about conserved charges, central charges and the behavior of the correlation functions of dual CFT by using the holographic renormalization method \cite{Alishahiha:2010bw,Kwon:2011jz}.

One purpose of this paper is an attempt to obtain mass and angular momentum of black holes using the first order formalism.  As we consider MMG theory, the field equation of this theory is composed of general TMG equation with a parameter $\gamma$ and an additional symmetric tensor $J_{\mu\nu}$ which comprises symmetric squared Shouten tensors in order to evade ``bulk vs. boundary clash''. The MMG field equation is given by
\begin{equation}\label{MMG-eq}
     \bar{\sigma} G_{\mu\nu} + \bar{\Lambda}_0 g_{\mu\nu} + \frac{1}{\mu} C_{\mu\nu} + \frac{\gamma}{\mu^2} J_{\mu\nu} = 0 \,,  
\end{equation}
with some shifted parameters $\bar{\sigma}$, $\bar{\Lambda}_0$ and a non-zero dimensionless parameter $\gamma$ as a function of parameter $\alpha$ \cite{Bergshoeff:2014pca}. Since parameters $\sigma$ and $\Lambda_0$ are no longer sign and cosmological constant respectively, they should be replaced by $\bar{\sigma}$ and $\bar{\Lambda}_0$. This equation cannot be obtained from an action for the metric alone by back-substituting the equation of the auxiliary fields $h$ into the MMG action. 
So dealing with field equations including auxiliary fields seems more correct and consistent way. Most of conserved charges are described by a surface integral having some tensors which are induced by the variation of a metric as its integrand. Using the first order formalism with field equations including auxiliary fields, we can obtain the charge variation which is represented by non-tensorial form structures as we can see below (\ref{Q-var}).
The conserved charges and central charges in MMG theory have been calculated by Tekin using ADT method \cite{Tekin:2014jna}.
The entropy of BTZ black hole in MMG theory have been obtained by Setare and Adami \cite{Setare:2015pva} with the first order formalism according to Tachikawa's method \cite{Tachikawa:2006sz}. 
Also the conserved charges in generalized minimal massive gravity (GMMG) theory \cite{Bergshoeff:2014bia,Setare:2014zea} have been obtained by Setare and Adami with the same method \cite{Setare:2015vea}. 
There were some studies of the properties of the linearized equation and holographic renormalization in MMG theory \cite{Alishahiha:2014dma,Alishahiha:2015whv}. 
Nowadays there was a study about black hole entropy as the horizon Noether charge for diffeomorphism and local Lorentz symmetry \cite{Jacobson:2015uqa}.
It has also been performed some studies for the quasi-local conserved charges by considering the Lorentz diffeomorphism invariant gravity theories
\cite{Setare:2015nla,Setare:2015cvv,Setare:2015gss,Setare:2016hmn}.

In this paper we use the Wald's method to obtain mass and angular momentum of black holes with the first order formalism. In this method mass and angular momentum is just defined by an integration of the variational form at spatial infinity (\ref{Mass-Ang}). Even though it is not an exact derivation of the charge variation form with the first order formalism, it is enough to get mass and angular momentum of the three dimensional gravity theories. This method seems to work well in cases of having some boundaries at infinity such as asymptotically Minkowski and $AdS$ spacetime. We consider well-known three dimensional gravity theories and deal with field equations including auxiliary fields using the first order formalism. We calculate well-known results of mass and angular momentum of black holes as some examples. Next we calculate mass and angular momentum of BTZ black hole and new type black hole of MMG theory as a new result.

Another purpose of this paper is to find the thermodynamic relation of new type black hole in MMG theory. The action of the three dimensional gravity theories can be represented by an integral of a Lagrangian three-form $L$ constructed as a sum of the wedge products of $N$ ``flavors" of Lorentz vector valued one-form fields, $\{ a^r | r=1,\cdots,N \}$, such as the dreibein $e^a$, the spin-connection $\omega^a$ and some proper auxiliary fields $h^a, f^a$ etc. to get a set of equivalent first-order equations. In this formulation the ``Chern-Simons-like" Lagrangian takes the form \cite{Bergshoeff:2014bia,Merbis:2014vja} 
\begin{equation}\label{CSL}
    L_{CSL} = \frac{1}{2} g_{rs} a^r \cdot da^s + \frac{1}{6} f_{rst} a^r \cdot (a^s \times a^t) \,,
\end{equation}
where $g_{rs}$ is a symmetric invertible constant flavor metric and $f_{rst}$ defines a totally symmetric coupling constants on the flavor space. The dot and cross 
represent the wedge product including the contraction of Lorentz vectors with $\eta_{ab}$ and $\epsilon_{abc}$ respectively. In order to get the boundary 
central charges of new type black hole in MMG theory, we investigate the Poisson brackets of the primary constraint functions by using the Hamiltonian analysis \cite{Bergshoeff:2014bia,Bergshoeff:2014pca}. Furthermore, we calculate the entropy of new type black hole using the charge variation form including 
the gravitational Chern-Simons term \cite{Tachikawa:2006sz} with the first order formalism. According to the prescription of the $AdS/CFT$, the entropy of a black hole can be interpreted in terms of the quantities of the dual CFT side through the Cardy formula \cite{Cardy:1986ie,Carlip:1998qw}. It is also well known that the conserved charges of the bulk gravity are related with the energies ${\cal E}_L$ and ${\cal E}_R$ of the dual CFT which is represented in terms of left and right moving central charges and square of temperatures. Comparing the entropy and mass of new type black hole with those of the Cardy formula and the charge relations in accordance with $AdS/CFT$ correspondence, we should suggest that the left and right moving temperatures of the dual CFT are equal to the Hawking temperature of new type black hole.

This paper is organized as follows. In section 2, we briefly review about Wald's method to define entropy, mass and angular momentum of a black hole 
\cite{Wald:1993nt,Jacobson:1993vj,Iyer:1994ys,Wald:1999wa}. In section 3, as some examples we compute masses and angular momenta of some black holes in TMG and NMG theories with the charge variation form obtained by using the first order formalism. In section 4, we obtain mass and angular momentum of BTZ black hole and mass of new type black hole in MMG theory as some new results . In section 5, we calculate the central charges and entropy of new type black hole. From these results, we obtain Smarr relation between mass and entropy of new type black hole. In section 6, We summarize our results and add some comments. Appendices are attached to the last to explain some useful formulae to calculate mass and angular momentum of the warped $AdS$ black hole in TMG theory.

\section{Brief review of the Wald formalism}

In this section we briefly survey the Wald formalism which has established entropy, mass and angular momentum of black holes. 
According to this formalism, black hole entropy is the integral of the diffeomorphism Noether charge associated with the horizon-generating
Killing field which vanishes on the Killing horizon \cite{Wald:1993nt}. If there exists a black hole solution with a Killing vector $\xi$ which generates 
a local symmetry of the solution, then the corresponding canonical mass and angular momentum of the solution are well defined at spatial infinity. 

Consider  a diffeomorphism invariant theory defined by a Lagrangian $n$-form $L$, where $n$ is the spacetime dimension. 
The variation $\delta L$ is induced by a field variation $\delta \phi$
\begin{equation}\label{varL}
    \delta L = E_{\phi} \delta \phi + d\Theta (\phi, \delta \phi)  \,.
 \end{equation}
where $\phi$ means the dynamical fields. The $E_{\phi}$ describes the field equation $E_{\phi}=0$ which is constructed from the dynamical variables 
$\phi$ and their first derivatives, and $(n-1)$-form $\Theta$ is the symplectic potential which 
is constructed by the dynamical fields and their first variations.
The $(n-1)$-form symplectic current is defined by the anti-symmetrized field variation of $\Theta$,
\begin{equation} \label{symp-cur}
  \omega(\phi, \delta \phi_1, \delta \phi_2) = \delta_1 \Theta (\phi, \delta_2 \phi) - \delta_2 \Theta (\phi, \delta_1 \phi)  \,.
\end{equation}
Then the symplectic form can be defined by an integration over a Cauchy surface $\Sigma$ in globally hyperbolic spacetime 
\begin{equation} \label{symp-form}
 \Omega (\phi, \delta_1 \phi, \delta_2 \phi) = \int_{\Sigma} \omega (\phi, \delta_1 \phi, \delta_2 \phi) , 
\end{equation}
where this surface is the field configuration space with unperturbed solution. 

Let $\xi$ be a vector field on spacetime manifold and consider the variation induced by a diffeomorphism generated by a vector field $\xi$,
\begin{equation}
  \delta_{\xi} \phi = \pounds_{\xi} \phi  \,.
\end{equation}
Then diffeomorphism invariant Lagrangian implies that the variation of the Lagrangian is equal to the Lie derivative of the Lagrangian under 
this variation,
\begin{eqnarray}
    \delta_{\xi} L = \pounds_{\xi} L = d i_{\xi} L \,.
\end{eqnarray}
Because the above equation is a total derivative, it shows that the vector fields $\xi$ on a spacetime generate infinitesimal local symmetries. 
According to the equation (\ref{varL}) under this variation a Noether current $(n-1)$-form $J_{\xi}$, which is defined by 
\begin{equation}\label{Ncurrent}
    J_{\xi} = \Theta(\phi, \pounds_{\xi} \phi) - i_{\xi} L \,,
\end{equation}
can be associated to each vector $\xi^a$. Applying exterior derivative to this current gives
\begin{equation}
 dJ_{\xi} = d\Theta(\phi, \pounds_{\xi} \phi) - di_{\xi} L = - E_{\phi} \pounds_{\xi} \phi  \,.
\end{equation}
Therefore the current $J_{\xi}$ is closed when equations of motion are satisfied, i.e. $E_{\phi}=0$. It means that the Noether current can be 
represented by the exact form,
\begin{equation}
   J_{\xi} = dQ_{\xi} \,,
\end{equation}
where $(n-2)$-form $Q_{\xi}$ is constructed from the fields and derivatives that are appearing in Lagrangian with $\xi$. 

In order to derive the first law of black hole mechanics for the perturbations of a black hole in an arbitrary diffeormorphism covariant theory, we investigate an identity at first. Consider $\phi$ to be any solution of the equations of motion. Let $\delta \phi$ be an arbitrary variation of the dynamical field off the solution $\phi$. Then we survey the variation of the Noether current
\begin{equation} \label{var-J}
      \delta J_{\xi} = \delta \Theta(\phi, \pounds_{\xi} \phi) - i_{\xi} \delta L \,.
\end{equation}    
Here we put $\xi$ to be an arbitrary fixed vector field in this variation, i.e. $\delta \xi = 0$. With (\ref{varL}), 
\begin{equation*}
       i_{\xi} \delta L = i_{\xi} (E_{\phi} \delta \phi + d \Theta (\phi, \delta \phi)) = \pounds_{\xi} \Theta(\phi, \delta \phi) - d i_{\xi} \Theta (\phi, \delta \phi) \,,
\end{equation*}
where we apply the equations of motion, $E_{\phi} = 0$. Therefore the relation (\ref{var-J}) becomes
\begin{equation}
     \delta J_{\xi} = \delta \Theta (\phi, \pounds_{\xi} \phi) - \pounds_{\xi} \Theta (\phi, \delta \phi) + di_{\xi} \Theta (\phi, \delta \phi)  \,.
\end{equation}
The phase space is the space of solutions to the field equation in the covariant framework. The variation $\delta_{\xi} \phi$ satisfying equations of motion describes the flow vector of the phase space corresponding to the 1-parameter family of diffeomorphisms generated by $\xi$. Then the variation of the Hamiltonian $H_{\xi}$ conjugate to $\xi$ is related to the symplectic form (\ref{symp-form}) 
\begin{equation} \label{var-H}
 \delta H_{\xi} = \int_{\Sigma} \omega(\phi, \delta \phi, \pounds_{\xi} \phi) \,,
\end{equation}
where $\Sigma$ is a Cauchy surface. If $\xi$ is a symmetry of all dynamical fields, i.e. $\pounds_{\xi} \phi = 0$, and their variation $\delta \phi$ satisfy the linearized equation. then the symplectic current is given by
\begin{eqnarray*}
     \omega (\phi, \delta \phi, \pounds_{\xi} \phi) &=& \delta \Theta(\phi, \pounds_{\xi} \phi) - \pounds_{\xi} \Theta (\phi, \delta \phi)    \\
                 &=& \delta J_{\xi} - d i_{\xi} \Theta(\phi, \delta \phi) = \delta dQ_{\xi} - di_{\xi} \Theta \,.
\end{eqnarray*}
Substituting the above formula into (\ref{var-H}) then the variation of the Hamiltonian becomes
\begin{equation}
     \delta H_{\xi} = \int_{\Sigma} \delta dQ_{\xi} - di_{\xi} \Theta = \oint_{\partial \Sigma} \delta Q_{\xi} - i_{\xi} \Theta \,,   \label{var-H2}
\end{equation}
where the integral over $\partial \Sigma$. Because of $\pounds_{\xi} \phi=0$ the symplectic current vanishes. So, eq.(\ref{var-H}) implies $\delta H_{\xi} = 0$.
Therefore the last line of the above formula becomes
\begin{equation}
     0 = \oint_{\partial \Sigma} \delta Q_{\xi} - i_{\xi} \Theta.
\end{equation}
Now consider a stationary black hole solution with a Killing field $\xi$ which generates a Killing horizon and vanishes on a bifurcation surface $\cal H$. 
If we choose the hypersurface $\Sigma$ to have its outer boundary
at spatial infinity and interior boundary at $\cal H$, then the variational identity can be expressed with two boundary terms
\begin{equation}\label{Hami}
  \int_{\cal H} \delta Q_{\xi} = \int_{\infty} \delta Q_{\xi} - i_{\xi} \Theta \,.
\end{equation}

If we assume that the asymptotic symmetries have been specified by the time translational Killing field and axial rotational one with the horizon angular velocity 
$\Omega_{\cal H}$, i.e.
\begin{equation}
    \xi = \frac{\partial}{\partial t} + \Omega_{\cal H} \frac{\partial}{\partial \phi}  \,,
\end{equation}
then the outer boundary integral of (\ref{Hami}) can be defined as the total energy and the angular momentum. Comparing (\ref{Hami}) with the first law of thermodynamics
\begin{equation}\label{1st-law}
 T_H \delta {\cal S} = \delta {\cal E} - \Omega_{\cal H} \delta {\cal J} \,,
\end{equation}
the left hand side gives the black hole entropy as the form 
\begin{equation}\label{entropy}
  {\cal S}_{\rm ent} = \frac{2\pi}{\kappa} \int_{\cal H} Q_{\xi}  \,.
\end{equation}
If we re-express the charge variation as a form 
\begin{equation} \label{Q-var}
     \delta \chi_{\xi} = \delta Q_{\xi} - i_{\xi} \Theta \,,
\end{equation}
the right hand side of (\ref{Hami}) gives the suitable definition of the total energy and the angular momentum up to constant, i.e. 
\begin{equation}
     \delta {\cal E} = \int_{\infty} \delta \chi_{\xi} \Big[ \frac{\partial}{\partial t} \Big] \,, \quad 
           \delta {\cal J} = - \int_{\infty} \delta \chi_{\xi} \Big[ \frac{\partial}{\partial \phi} \Big] \,.
\end{equation}
In this paper we intend to deal with three dimensional gravity theories. 
So we define the variation of the mass and angular momentum of a black hole in three dimensional gravity theories as the form
\begin{equation} \label{Mass-Ang}
   \delta {\cal M} = - \frac{1}{8\pi G} \int_{\infty} \delta \chi_{\xi} \Big[ \frac{\partial}{\partial t} \Big]  \,,   \quad
        \delta {\cal J} =  \frac{1}{8\pi G} \int_{\infty} \delta \chi_{\xi} \Big[ \frac{\partial}{\partial \phi} \Big] \,,
\end{equation}
where $1/8\pi G$ is a constant for three dimensional gravity theories.
%

\section{Masses and angular momenta of black holes in three dimensional gravity theories : examples}

\subsection{Topologically massive gravity}

We consider diffeomorphism invariant Lagrangians in three dimensional gravity theories with the first order orthonormal frame.
In these cases the Lagrangian can be written in terms of  local Lorentz vector-valued 1-form frame fields $e^a$ and 
connection 1-forms $\omega^a_{~b}$. The spacetime metric tensor is denoted by the relation
\begin{equation}
   g_{\mu \nu} = \eta_{ab} e^a_{\mu} e^b_{\nu} \,,
\end{equation}
where $\eta_{ab}$ is the Minkowski metric. The connection 1-form $\omega^a_{~b}$ can be expressed by a dualised form $\omega^a$ as 
\begin{equation}
    \omega_a = \frac{1}{2} \epsilon_{abc} \omega^{bc} \,.
\end{equation}
Both $e^a$ and $\omega^a$ are considered as independent variables to be varied separately in the action.  
The action is represented by the integral of the Lagrangian 3-form $L$ which can be constructed from wedge products of 
the frame fields $e^a$ and connections $\omega^a$. Firstly we consider 3-dimensional gravitational Chern-Simons theory or Topologically Massive
Gravity theory (TMG).
The Lagrangian form of this theory (TMG) is given by
\begin{equation}
  L = - \sigma e\cdot R + \frac{\Lambda_0}{6} e \cdot e \times e + \frac{1}{2\mu} \big( \omega \cdot d\omega 
        + \frac{1}{3}\omega \cdot \omega \times \omega \big) + h \cdot T(\omega) \,,
\end{equation}
where $\Lambda_0$ is a cosmological constant and $\sigma$ is a sign. This action is constructed with three Lorentz vector-valued 1-forms 
$(e, \omega, h)$ and local Lorentz covariant torsion $T(\omega)$ and curvature 2-form $R(\omega)$ which are defined by
\begin{equation} \label{T-R}
      T(\omega) = De = de + \omega \times e \,, \quad  R(\omega) = d\omega + \frac{1}{2} \omega \times \omega \,.
\end{equation}
In the action Lorentz indices $a, b, c, \cdots$ are suppressed, and contractions of $\eta_{ab}$ and $\epsilon_{abc}$ with wedge products are represented by 
the sign `$\cdot$' and `$\times$' respectively. The third term of the action with the factor $1/\mu$ describes the `Local Lorentz Chern-Simons' term.
The auxiliary field $h$ is a Lagrange multiplier for the torsion-free constraint and has the same parity and dimension of $\omega$.   

The variation of the action is given by
\begin{eqnarray}
   \delta L &=& \delta e \cdot \Big( - \sigma R + \frac{\Lambda_0}{2} e \times e + D h \Big) 
                        + \delta \omega \cdot \Big( -\sigma D e + \frac{1}{\mu} R + e \times h \Big) + \delta h \cdot T(\omega)   \nonumber   \\
                && + d \Big( -\sigma \delta \omega \cdot e + \frac{1}{2\mu} \delta \omega \cdot \omega + \delta e \cdot h \Big)   
                      = E_{\phi} \delta \phi + d\Theta  \,,   
\end{eqnarray}
where $D$ is the Lorentz covariant exterior derivative. 
From the above variation we obtain equations of motion as follows
\begin{equation}\label{eoms} 
    - \sigma R + \frac{\Lambda_0}{2} e \times e + D h = 0 \,,  ~~~
    -\sigma De + \frac{1}{\mu}R + e \times h = 0 \,,    ~~~
   T(\omega) = De = 0  \,,    
\end{equation}
and symplectic potential
\begin{equation}
   \Theta = - \sigma \delta \omega \cdot e + \frac{1}{2\mu} \delta \omega \cdot \omega + \delta e \cdot h \,.
\end{equation}
Following the Wald's formalism, we can find Noether current using (\ref{Ncurrent}),
\begin{equation}
    j_{\xi} = dQ_{\xi} = \Theta (\phi, \pounds_{\xi} \phi) - i_{\xi} L     
              = d \big( - \sigma i_{\xi} \omega \cdot e + \frac{1}{2\mu} i_{\xi} \omega \cdot \omega + i_{\xi} e \cdot h \big) \,. 
\end{equation}
So, from the above equation we can read the Noether charge
\begin{equation}
   Q_{\xi} =  - \sigma i_{\xi} \omega \cdot e + \frac{1}{2\mu} i_{\xi} \omega \cdot \omega + i_{\xi} e \cdot h \,.
\end{equation}
From (\ref{Hami}) we can calculate the following variation form
\begin{equation}\label{chi-var} 
   \delta \chi_{\xi} = \delta Q_{\xi} - i_{\xi} \Theta     
            = - \sigma (i_{\xi} \omega \cdot \delta e + \delta \omega \cdot i_{\xi} e)  
                + \frac{1}{\mu} i_{\xi} \omega \cdot \delta \omega + i_{\xi} e \cdot \delta h + \delta e \cdot i_{\xi} h \,.    
\end{equation}
Now we consider a general metric form 
\begin{equation}\label{gen-metric}
    ds^2 = - f(r)^2 dt^2 + \frac{dr^2}{f(r)^2} + r^2 (d\phi + N(r) dt)^2  \,.
\end{equation}
In order to apply the first order orthonormal frame formalism to this theory we need to take 1-form frame fields as follows,
\begin{equation} \label{dreibein}
     e^0 = f(r) dt \,, ~  e^1 = \frac{dr}{f(r)} \,, ~  e^2 = r(d\phi + N(r) dt) \,.
\end{equation}
We firstly consider the third equation of (\ref{eoms}) which means the torsion free condition in (\ref{T-R}). 
For convenience we express functions $f(r)$ and $N(r)$ as the abbreviated form without $r$.  
Then we can find connection 1-forms $\omega^a$,
\begin{eqnarray}\label{con-1form}
    && \omega^0 = \frac{1}{2} r N' e^0 + \frac{f}{r} e^2 \,,  \nonumber   \\
    && \omega^1 = \frac{1}{2} r N' e^1 \,,      \\
    && \omega^2 = - \frac{1}{2} r N' e^2 + f' e^0  \,,   \nonumber 
\end{eqnarray}
where ` $\prime$ ' denotes the differentiation of a function with respect to $r$. It is easy to find curvature 2-forms and auxiliary field $h^a$ 
by substituting (\ref{dreibein}), (\ref{con-1form}) into the definition of curvature 2-form (\ref{T-R}) and second equation of (\ref{eoms}).

%
%
%
%
For the first example we now consider BTZ black hole. The metric of BTZ black hole with $\Lambda_0 = - \frac{1}{\ell^2}$ is given by 
the metric form (\ref{gen-metric}) with functions
\begin{equation} \label{BTZ}
  f(r) = \frac{\sqrt{(r^2 - r_+^2)(r^2 - r_-^2)}}{\ell r} \,, \quad  N(r) = - \frac{r_+ r_-}{\ell r^2}  \,.
\end{equation}
%
%
%
%
%
Using the above functions (\ref{BTZ}), the auxiliary fields $h^a$ can be simply expressed as
\begin{equation}\label{aux-BTZ}
           h^a = \frac{1}{2\mu \ell^2} e^a \,.
\end{equation}
Now in order to compute mass and angular momentum of a black hole we consider the charge variation (\ref{Q-var}).  
Using the above formula (\ref{aux-BTZ}), we can rephrase the charge variation form (\ref{chi-var}) as follows 
\begin{equation} \label{var-Chi}
      \delta \chi_{\xi} = -\sigma i_{\xi} \omega \cdot \delta e - \sigma \delta \omega \cdot i_{\xi} e  
                   + \frac{1}{\mu} i_{\xi} \omega \cdot \delta \omega + \frac{1}{\mu \ell^2} i_{\xi} e \cdot \delta e  \,. 
\end{equation}

The mass and angular momentum of a black hole is defined by (\ref{Mass-Ang}) on the boundary, i.e. spatial infinity. 
To compute the charge variation form for mass and angular momentum
 of a black hole we examine the interior products and variations of frame fields and connection 1-forms. 
%
%
%
%
%
Since we are now dealing with the variation of Hamiltonian (\ref{var-H2}) at spatial boundaries on the Cauchy surface, 
 we only need to consider $d\phi$ component to compute the charge variation form $\delta \chi_{\xi}$. 
 The variation forms of connection 1-forms to be related to $d\phi$ component are given by
 \begin{equation}\label{BTZ-var}
      \delta \omega^0 = \delta f d\phi \,, \quad  \delta \omega^2 = - \frac{1}{2} r^2 \delta N' d\phi \,.
 \end{equation}
There are no variation forms of the frame fields related to $d\phi$ component. 
Therefore the charge variation for a black hole mass in TMG becomes
\begin{equation}\label{chi-BTZ-MT}
      \delta \chi_{\xi} \Big[ \frac{\partial}{\partial t} \Big]
               = \sigma \Big( f \delta f + \frac{1}{2} r^3 N \delta N' \Big) d\phi       
                 - \frac{1}{\mu} \bigg\{ f\Big( N + \frac{1}{2} r N' \Big) \delta f         
                           + \frac{1}{2} r^2 \Big( ff' - \frac{1}{2} r^2 N N' \Big) \delta N' \bigg\} d\phi   \,.     
\end{equation}
The variation of frame fields $e^a$ and connections $\omega^a$ should
be performed by the coordinate of the horizons $r_+$ and $r_-$ since these horizon coordinates behave as physical quantities.
In other words, variation means the difference between solutions and background.
%
%
%
Then from the definition of (\ref{Mass-Ang}) we can get the variation formula for the mass,
\begin{equation}
     \delta {\cal M} = - \frac{1}{8\pi G} \int_{\infty} \delta \chi_{\xi} \Big[ \frac{\partial}{\partial t} \Big]      
                  = \frac{1}{4G} \bigg\{ \frac{\sigma}{\ell^2} (r_+ \delta r_+ + r_- \delta r_-)      
                       + \frac{1}{\mu \ell^3} (r_+ \delta r_- + r_- \delta r_+) \bigg\}   \,.          
\end{equation}
Integrating and considering the total variation of the right hand side of the above formula, we can get the mass of BTZ black hole in TMG theory,
\begin{equation}
      {\cal M} = \sigma \frac{r_+^2 + r_-^2}{8G \ell^2} + \frac{r_+ r_-}{4G \mu \ell^3}   \,,
\end{equation} 
which is the same result in \cite{Bouchareb:2007yx,Hotta:2008yq}.
In order to get the angular momentum we need to consider the asymptotic rotational symmetry, i.e. taking the Killing vector as
$\xi = \frac{\partial}{\partial \phi}$. Following the same procedure with the case of the mass the charge variation (\ref{var-Chi}) with this rotation Killing vector at spatial infinity is given by
\begin{equation} \label{CJ-BTZ}
    \delta \chi_{\xi} \Big[ \frac{\partial}{\partial \phi} \Big] 
                 = \bigg\{ \sigma \frac{1}{2} r^3 \delta N' + \frac{1}{\mu} \Big( -f \delta f + \frac{1}{4} r^4 N' \delta N' \Big) \bigg\} d\phi \,.
\end{equation}
Substituting this form into the definition (\ref{Mass-Ang}) for the angular momentum formula, then we obtain
\begin{equation}
    \delta {\cal J} = \frac{1}{8 \pi G} \int_{\infty} \delta \chi_{\xi} \Big[ \frac{\partial}{\partial \phi} \Big]      
            = \frac{1}{4G} \bigg\{ \frac{\sigma}{\ell} (r_+ \delta r_- + r_- \delta r_+)       
                      + \frac{1}{\mu \ell^2} (r_+ \delta r_+ + r_- \delta r_-) \bigg\}  \,.       
\end{equation}
Performing the integral and considering the total variation of the above formula, the angular momentum of BTZ black hole is given by
\begin{equation} \label{Ang-BTZ}
      {\cal J} = \sigma \frac{r_+ r_-}{4G \ell} + \frac{r_+^2 + r_-^2}{8G \mu \ell^2} \,.
\end{equation}
These results are also the same with \cite{Bouchareb:2007yx,Hotta:2008yq}.
The charge variation (\ref{CJ-BTZ}) can be re-expressed by a total variation form. So, we can represent the charge form $\chi_{\xi}$ as 
\begin{equation} \label{CJ-BTZ-2}
    \chi_{\xi} \Big[ \frac{\partial}{\partial \phi} \Big] 
            = \bigg\{ \sigma \frac{1}{2} r^3 N' + \frac{1}{2\mu} \Big( - f^2 + \frac{1}{4} r^4 N'^2 \Big) \bigg\} d\phi \,.
\end{equation}
This value should be computed at spatial infinity. The part of this value with negative power of $r$ vanishes as $r$ goes to infinity. 
$r^2$ term does not give any contribution to (\ref{CJ-BTZ}) because there is no $r_+$ and $r_-$ in this coefficient and this term only means AdS background. 
So, considering the constant term is enough to compute the charge form $\chi_{\xi}$. 
Then we can obtain the same result (\ref{Ang-BTZ}) with the charge definition (\ref{Mass-Ang}) and (\ref{CJ-BTZ-2}). 
Now we investigate the space-like warped $AdS_3$ black hole solution as the second example \cite{Anninos:2008fx,Moussa:2003fc}.
The metric of the space-like warped black hole solution is given by
\begin{equation} \label{warped}
      ds^2 = - N(r)^2 dt^2 + \frac{\ell^4 dr^2}{4 R(r)^2 N(r)^2} + \ell^2 R(r)^2 ( d\theta + N^{\theta}(r) dt )^2  \,,  
\end{equation}
where
\begin{eqnarray} \label{wap-func}
      R(r)^2 &=& \frac{r}{4} \Big( 3(\nu^2-1) r + (\nu^2 + 3)(r_+ + r_-) - 4\nu \sqrt{r_+ r_- (\nu^2+3)} \Big)  \,,     \nonumber   \\
     N(r)^2 &=& \frac{\ell^2 (\nu^2+3) (r - r_+)(r - r_-)}{4 R(r)^2}  \,,   \\     
     N^{\theta}(r) &=& \frac{2\nu r - \sqrt{r_+ r_- (\nu^2 + 3)}}{2 R(r)^2}  \,.    \nonumber
\end{eqnarray}
From the metric we can easily read off the 1-form frame fields
\begin{equation} \label{drei-wap}
   e^0 = N(r) dt \,, ~ e^1 = \frac{\ell^2 dr}{2R(r) N(r)}  \,,  ~ e^2 = \ell R(r) (d\theta + N^{\theta}(r) dt) \,,
\end{equation}
and 1-form connections can be calculated using the third equation of (\ref{eoms}), i.e. torsion-free conditions, as follows 
\begin{eqnarray} \label{con-wap}
   &&   \omega^0 = \frac{R^2 N^{\theta \prime}}{\ell} e^0 + \frac{2NR'}{\ell^2} e^2  \,,    \nonumber   \\
   &&   \omega^1 = \frac{R^2 N^{\theta \prime}}{\ell} e^1   \,,     \\
   &&   \omega^2 = - \frac{R^2 N^{\theta \prime}}{\ell} e^2 + \frac{2RN'}{\ell^2} e^0  \,.   \nonumber
\end{eqnarray}
Concrete forms of curvature 2-forms and auxiliary fields for the space-like warped $AdS_3$ black hole are relegated to the appendix A.

In order to find the black hole mass we consider $\xi = \frac{\partial}{\partial t}$, then non-vanishing interior products and variations of the frame fields, connections and auxiliary fields can be represented by appendix B. Then the charge variation $\delta \chi_{\xi}$ for the mass of the warped $AdS_3$ black hole is given by 
\begin{eqnarray} \label{c1-var-wap}
     \delta \chi_{\xi} \Big[ \frac{\partial}{\partial t} \Big] 
          &=& - \sigma \bigg\{ \Big( - \frac{R^2 N^{\theta \prime}}{\ell} \ell R N^{\theta} + \frac{2RN'}{\ell^2} N \Big)  \ell \delta R     
                  - N \frac{2}{\ell} \delta(NRR')  - \ell R N^{\theta}  \delta(R^3 N^{\theta \prime}) \bigg\} d\theta   \nonumber  \\
          && - \frac{1}{\mu} \bigg\{ \Big( \frac{R^2 N^{\theta \prime}}{\ell} N + \frac{2NR'}{\ell^2} \ell R N^{\theta} \Big) \frac{2}{\ell} \delta(NRR')      \nonumber    \\
          && + \Big( - \frac{R^2 N^{\theta \prime}}{\ell} \ell R N^{\theta} + \frac{2RN'}{\ell^2} N \Big) \delta(R^3 N^{\theta \prime}) \bigg\} d\theta    \nonumber    \\
          && + \bigg[ \ell R N^{\theta} \bigg\{ \frac{\ell}{2\mu} \Big( \frac{\nu^2}{\ell^2}    
                + \frac{3(\nu^2 - 1)}{\ell^2} \Big) \delta R + \frac{\ell}{\mu} \delta(F R) \bigg\}         \nonumber    \\
          && + \bigg\{ \frac{1}{2\mu} \Big( \frac{\nu^2}{\ell^2} + \frac{3(\nu^2 - 1)}{\ell^2} + 2F \Big) \ell R N^{\theta}      
                + \frac{1}{\mu} G N \bigg\} \ell \delta R  + N \frac{\ell}{\mu} \delta(G R) \bigg]  d\theta  \,.     
\end{eqnarray}
In this theory, the mass of the black hole can be obtained using the formula
\begin{eqnarray*}
    \delta{\cal M} &=& - \frac{1}{8\pi G \ell} \int_{\infty} \delta \chi_{\xi} \Big[ \frac{\partial}{\partial t} \Big]      \\
                        &=& \frac{1}{8\pi G \ell} \int_{\infty} \frac{\ell (\nu^2 + 3)}{6} \bigg\{ \delta r_+ + \delta r_-                 
                            - \frac{1}{\nu} \frac{\sqrt{r_+ r_- (\nu^2 + 3)} }{2r_+ r_-} (r_- \delta r_+ + r_+ \delta r_-) \bigg\} d\theta \,,
\end{eqnarray*}
where we replace the gravitational constant $G$ with $G\ell$ in the definition (\ref{Mass-Ang}). 
The metric for the warped $AdS_3$ black hole (\ref{warped}) is written as the form with dimensionless coordinates. 
Only the cosmological constant $\ell$ has a length dimension. So, we should change these coordinates to be dimensionful to give the correct result. 
Therefore integrating the above formula we can get 
\begin{equation} \label{M-warped}
        {\cal M} = \frac{(\nu^2 + 3)}{24G} \Big( r_+ + r_- - \frac{1}{\nu} \sqrt{r_+ r_- (\nu^2 + 3)} \Big)   \,,
\end{equation}
which is the same result with \cite{Anninos:2008fx}.

To find the black hole angular momentum we need to consider $\xi = \frac{\partial}{\partial \theta}$. Using the appendix B, the charge variation is given by 
\begin{eqnarray} \label{c2-var-wap}
 \delta \chi_{\xi} \Big[ \frac{\partial}{\partial \theta} \Big] 
          &=& - \sigma \bigg\{ - \frac{R^2 N^{\theta \prime}}{\ell} \ell R \ell \delta R - \ell R \delta(R^3 N^{\theta \prime}) \bigg\} d\theta    \nonumber   \\
  && + \frac{1}{\mu} \bigg\{ - \frac{2N R'}{\ell^2} \ell R \frac{2}{\ell} \delta(NRR')   
                  + \frac{R^2 N^{\theta \prime}}{\ell} \ell R \delta(R^3 N^{\theta \prime}) \bigg\} d\theta      \nonumber   \\
  && + \bigg[ \ell R \bigg\{ \frac{\ell}{2\mu} \Big( \frac{\nu^2}{\ell^2} + \frac{3(\nu^2 - 1)}{\ell^2} \Big) \delta R 
                 + \frac{\ell}{\mu} \delta(F R) \bigg\}               \nonumber   \\
  && + \frac{1}{2\mu} \Big( \frac{\nu^2}{\ell^2} + \frac{3(\nu^2 - 1)}{\ell^2} + 2F \Big) \ell R \ell \delta R \bigg] d\theta  \,.
\end{eqnarray}
Since the above formula can be represented by the total variation form, we simplify this form to become a charge form 
\begin{equation} \label{Ang-C1}
  \chi_{\xi} \Big[ \frac{\partial}{\partial \theta} \Big]  
           = \bigg\{ \sigma \ell R^4 N^{\theta \prime} - \frac{2}{\mu \ell^2} (NRR')^2 + \frac{1}{2\mu} (R^3 N^{\theta \prime})^2  
              + \frac{\ell^2}{2\mu} \Big( \frac{\nu^2}{\ell^2} + \frac{3(\nu^2 - 1)}{\ell^2} \Big) R^2 + \frac{\ell^2}{\mu} R^2 F \bigg\} d\theta  \,. 
\end{equation}
When $r$ goes to infinity the charge form (\ref{Ang-C1}) can be expressed as a polynomial of $r$ with $r^2$ term as its highest order. 
The coefficients of $r^2$ and $r$ vanish to leave the constant term only.
Substituting all functions into the above form with $\sigma = 1, \frac{1}{\mu} = \frac{\ell}{3\nu}$, we obtain
\begin{equation}
    \chi_{\xi} \Big[ \frac{\partial}{\partial \theta} \Big] 
            = - \frac{\ell \nu (\nu^2 + 3)}{24} \bigg[ \Big( r_+ + r_- - \frac{1}{\nu} \sqrt{r_+ r_- (\nu^2 + 3)} \Big)^2    
                 - \frac{(5\nu^2 + 3)}{4\nu^2} (r_+ - r_-)^2 \bigg] d\theta  \,.        \label{Chi-Ang}
\end{equation}
From the definition of the angular momentum eq.(\ref{Mass-Ang}), we can get the angular momentum of the warped $AdS_3$ black hole
\begin{equation} \label{ang-warped}
    {\cal J} = - \frac{\ell \nu (\nu^2 + 3)}{96 G} \bigg[ \Big( r_+ + r_- - \frac{1}{\nu} \sqrt{r_+ r_- (\nu^2 + 3)} \Big)^2     
                     - \frac{(5\nu^2 + 3)}{4\nu^2} (r_+ - r_-)^2 \bigg]  \,.
\end{equation}
To obtain the above result we also consider the change of the coordinate to be dimensionful. So, we change the gravitational constant $G$ to $G\ell$. 
Because these coordinates changes the angular velocity's dimension, so in order to have correct dimension we need to change the angular velocity 
with $1/ \ell$ and rotational Killing vector with $\ell$. Then we can obtain the correct angular momentum (\ref{ang-warped}) 
which is the same result with \cite{Anninos:2008fx}.

\subsection{New Massive Gravity}

The Lagrangian of New Massive Gravity (NMG) theory can be represented by the first order form 
\begin{equation}
   L_{NMG} =  - \sigma e \cdot R + \frac{\Lambda_0}{6} e \cdot e \times e + h \cdot De - \frac{1}{m^2} f \cdot \big( R + \frac{1}{2} e \times f \big) \,,
\end{equation}
where $h$ and $f$ are auxiliary fields and $m$ is a mass parameter \cite{Hohm:2012vh,Blagojevic:2010ir}.
The variation of NMG Lagrangian is given by
\begin{eqnarray}\label{var-NMG}
   \delta L &=&  d \big( - \sigma \delta \omega \cdot e - \frac{1}{m^2} \delta \omega \cdot f + \delta e \cdot h \big)  
                       + \delta h \cdot De - \frac{1}{m^2} \delta f \cdot ( R + e \times f )     \nonumber    \\
                &&  + \delta \omega \cdot \Big( - \sigma De + h \times e - \frac{1}{m^2} Df \Big)       
                       + \delta e \cdot \Big( - \sigma R + \frac{\Lambda_0}{2} e \times e + Dh - \frac{1}{2m^2} f \times f \Big)     \nonumber   \\
                &=& E_{\phi} \delta \phi + d \Theta  \,.   
\end{eqnarray}
From the above variation we can obtain equations of motion as follows
\begin{eqnarray}\label{eom-NMG}
    &&  De = 0 \,, \quad  R + (e \times f) = 0 \,, \quad  Df - m^2 (e \times h) = 0 \,,  \nonumber  \\
    &&  Dh - \sigma R + \frac{\Lambda_0}{2} e \times e - \frac{1}{2m^2} f \times f = 0  \,.    
\end{eqnarray}
The symplectic potential of NMG Lagrangian can be read off from the variation (\ref{var-NMG})  
\begin{equation}
   \Theta =  - \sigma \delta \omega \cdot e - \frac{1}{m^2} \delta \omega \cdot f + \delta e \cdot h \,.
\end{equation}
Using the definition of the Noether current we can find the Noether charge as 
\begin{equation}
       Q_{\xi} =  -\sigma i_{\xi} \omega \cdot e - \frac{1}{m^2} i_{\xi} \omega \cdot f + i_{\xi} e \cdot h   \,.         
\end{equation}
From the above two results, i.e. symplectic potential and Noether charge, we can calculate the charge variation
\begin{equation} \label{chi-NMG}
    \delta \chi_{\xi} = - \sigma ( i_{\xi} \omega \cdot \delta e + i_{\xi} e \cdot \delta \omega )
                 - \frac{1}{m^2} ( i_{\xi} \omega \cdot \delta f + i_{\xi} f \cdot \delta \omega )     
                 + i_{\xi} e \cdot \delta h + i_{\xi} h \cdot \delta e   \,.
\end{equation}
Firstly, we investigate BTZ black hole solution in NMG theory.  BTZ black hole is a solution of NMG theory 
with a cosmological constant $\Lambda_0$ which appears (\ref{condi-NMG-BTZ}) below. 
So, we can use the same metric form with (\ref{gen-metric}), functions (\ref{BTZ}) and the same frame 1-form fields (\ref{dreibein}).
%
Using second and third equations of motion of (\ref{eom-NMG}), we can get two auxiliary fields
\begin{equation}\label{aux-NMG-BTZ}
   f^a = \frac{1}{2\ell^2} e^a \,, \quad  h^a = 0 \,.
\end{equation}
With auxiliary fields solution (\ref{aux-NMG-BTZ})  a parameter condition can be appeared by solving the fourth equation of (\ref{eom-NMG}) 
\begin{equation}\label{condi-NMG-BTZ}
    \frac{\sigma}{\ell^2} + \Lambda_0 - \frac{1}{4m^2 \ell^4} = 0 \,.
\end{equation}
We can re-express the symplectic potential $\Theta$ and the Noether charge $Q_{\xi}$ by inserting auxiliary fields (\ref{aux-NMG-BTZ}) into (\ref{chi-NMG})
and then calculate the charge variation as follows
\begin{equation}
   \delta \chi_{\xi} = - \Big( \sigma + \frac{1}{2m^2 \ell^2} \Big) ( i_{\xi} \omega \cdot \delta e + i_{\xi} e \cdot \delta \omega ) \,.
\end{equation}
Performing the interior products and variations of frame 1-form fields and connection 1-forms with $\xi = \frac{\partial}{\partial t}$ we obtain
\begin{equation}
      \delta \chi_{\xi} \Big[ \frac{\partial}{\partial t} \Big] 
                 = \Big( \sigma + \frac{1}{2m^2 \ell^2} \Big) \Big( f \delta f + \frac{1}{2} r^3 N \delta N' \Big) d\phi    \,, 
\end{equation} 
as $r$ goes to infinity. From the definition of the black hole mass (\ref{Mass-Ang}) we can get 
\begin{equation*}
    \delta {\cal M} = - \frac{1}{8\pi G} \int_{\infty} \delta \chi_{\xi} \Big[ \frac{\partial}{\partial t} \Big]       
            = \frac{1}{4G \ell^2} \Big( \sigma + \frac{1}{2m^2 \ell^2} \Big) \big( r_+ \delta r_+ + r_- \delta r_- \big)  \,.
\end{equation*}
Therefore we can obtain the mass of BTZ black hole in NMG theory,
\begin{equation}
        {\cal M} = \frac{r_+^2 + r_-^2}{8G\ell^2} \Big( \sigma + \frac{1}{2m^2 \ell^2} \Big)    \,.
\end{equation}
To find the angular momentum we consider the Killing vector $\xi = \frac{\partial}{\partial \phi}$. 
Then as $r$ goes to infinity, the charge variation becomes 
\begin{equation}
    \delta \chi_{\xi} \Big[ \frac{\partial}{\partial \phi} \Big]
                  = \Big( \sigma + \frac{1}{2m^2 \ell^2} \Big) \cdot \frac{1}{2} r^3 \delta N' d\phi    \,.
\end{equation}
Applying the above formula to the definition of the angular momentum (\ref{Mass-Ang}), we can get 
\begin{equation*}
     \delta {\cal J} = \frac{1}{8\pi G} \int_{\infty} \delta \chi_{\xi} \Big[ \frac{\partial}{\partial \phi} \Big]       
               = \frac{1}{4G \ell} \Big( \sigma + \frac{1}{2m^2 \ell^2} \Big) (r_- \delta r_+ + r_+ \delta r_-) \,.
\end{equation*}
Therefore we can obtain the angular momentum of the BTZ black hole in NMG theory,
\begin{equation}
    {\cal J} = \frac{r_+ r_-}{4G\ell}  \Big( \sigma + \frac{1}{2m^2 \ell^2} \Big)   \,.
\end{equation}

Now we investigate the new type black hole solution which appears as another solution in NMG theory \cite{Bergshoeff:2009aq}.  
The metric form of this black hole is given by
\begin{equation}\label{new-BH}
        ds^2 = - f(r)^2 dt^2 + \frac{dr^2}{f(r)^2} + r^2 d\phi^2   \,,   
\end{equation}
where
\begin{equation}\label{f-G}
    f(r) = \frac{\sqrt{(r - r_+)(r - r_-)}}{\ell}   \,.
\end{equation}
This non-rotating new type black hole solution is represented by the general form (\ref{gen-metric}) with $N(r) = 0$. 
Solving equations of motion in (\ref{eom-NMG}), then we can find 
\begin{equation}\label{1f-NMG}
       e^0 = f(r) dt \,, \quad  e^1 = \frac{dr}{f(r)}  \,, \quad  e^2 = r d\phi \,,
\end{equation}
\begin{equation}\label{nt-con}
     \omega^0 = \frac{f}{r} e^2 \,, \quad  \omega^1 = 0 \,, \quad  \omega^2 = f' e^0 \,.
\end{equation}
%
%
%
Solving the second and third equations of motion in (\ref{eom-NMG}) with (\ref{f-G}), we can simply determine the auxiliary fields $f^a$ 
\begin{equation}\label{nt-aux-com}
       f^0 = \frac{1}{2 \ell^2} e^0 \,,  ~  f^1 = \frac{1}{2 \ell^2} e^1 \,, ~  f^2 = \frac{1}{2 \ell^2} \Big( 1 - \frac{r_+ + r_-}{r} \Big) e^2 \,,
\end{equation}
%
%
and all auxiliary fields $h^a$ vanish.
%

%
So, the charge variation (\ref{chi-NMG}) with the condition, $h^a = 0$, becomes 
\begin{equation}\label{chi-nt}
    \delta \chi_{\xi} = - \sigma ( i_{\xi} \omega \cdot \delta e + i_{\xi} e \cdot \delta \omega )
                 - \frac{1}{m^2} ( i_{\xi} \omega \cdot \delta f + i_{\xi} f \cdot \delta \omega )   \,.
\end{equation}
%
%
For the computation of the charge variation we need to get the variations of frame fields, connection 1-forms and auxiliary fields.
The non-vanishing $d\phi$ components of useful variations are given by
\begin{equation}\label{var-nt}
    \delta \omega^0 = \delta f d\phi \,, \quad  \delta f^2 = \delta (ff') d\phi \,.
\end{equation}
%
%
%
All variations of the frame fields related to $d\phi$ component vanish. 
Because we are dealing with the non-rotating new type black hole, we only consider the computation of the black hole mass.
%
%
Performing the calculation of (\ref{chi-nt}) for the Killing vector $\xi= \frac{\partial}{\partial t}$ with the function (\ref{f-G}),  
we can get the charge variation 
\begin{eqnarray*}
     \delta \chi_{\xi} \Big[ \frac{\partial}{\partial t} \Big] &=&  - \frac{1}{2 \ell^2} \Big( \sigma + \frac{1}{2m^2 \ell^2} \Big) 
                   \cdot  \big[  (r - r_-) \delta r_+ +(r - r_+)  \delta r_-  \big] d\phi      \\
         && + \frac{1}{4 m^2 \ell^4} ( 2 r - r_+ - r_-) ( \delta r_+ + \delta r_- )  d\phi   \,. 
\end{eqnarray*}
Solving equations of motion in (\ref{eom-NMG}) we can get two parameter conditions
\begin{equation}
    \frac{\sigma}{\ell^2} + \Lambda_0 - \frac{1}{4m^2 \ell^4} = 0 \,,  \quad  - \frac{\sigma}{2\ell^2} + \frac{1}{4m^2 \ell^4} = 0 \,.
\end{equation} 
From above conditions we should take a relation $\sigma = 2 m^2 \ell^2 = 1$, then the other condition gives a relation $\Lambda_0 = - \frac{1}{2\ell^2}$.
So we can rearrange the charge variation such as
\begin{equation}
         \delta \chi_{\xi} \Big[ \frac{\partial}{\partial t} \Big] = - \frac{1}{2 \ell^2}  ( r_+ - r_- ) ( \delta r_+ - \delta r_- )  d\phi  \,.
\end{equation}
Then the variation of the mass of the black hole with the definition (\ref{Mass-Ang}) is given by
\begin{equation}
   \delta {\cal M}  = - \frac{1}{8\pi G} \int_{\infty} \delta \chi_{\xi} \Big[ \frac{\partial}{\partial t} \Big]
                    = \frac{1}{8G \ell^2} ( r_+ - r_- ) ( \delta r_+ -  \delta r_- ) \,.
\end{equation}
So we can obtain the mass of new type black hole 
\begin{equation}
    {\cal M} = \frac{1}{16 G \ell^2} ( r_+ - r_- )^2  \,.
\end{equation}
This result has been computed in \cite{Nam:2010dd,Nam:2010ub,Kwon:2011jz}.

\section{ Masses and Angular momenta of black holes in Minimal Massive Gravity theory}

The Lagrangian of Minimal Massive Gravity (MMG) theory is given by
\begin{eqnarray}\label{MMG-action}
   L_{\rm MMG} &=& L_{\rm TMG} + \frac{\alpha}{2} e \cdot h \times h     \nonumber  \\
                  &=& - \sigma e \cdot R + \frac{\Lambda_0}{6} e \cdot e \times e + h \cdot T(\omega)    
                          + \frac{1}{2\mu} \Big( \omega \cdot d \omega + \frac{1}{3} \omega \cdot \omega \times \omega \Big)
                          + \frac{\alpha}{2} e \cdot h \times h \,,  
\end{eqnarray}
where the gravitational Chern-Simons term and some additional term with auxiliary fields $h^a$ are included \cite{Bergshoeff:2014pca}.
In order to get equations of motion we investigate the variation of the MMG Lagrangian, then the variation becomes 
\begin{eqnarray*}
     \delta L_{MMG} &=& \delta L_{TMG} + \frac{\alpha}{2} \delta ( e \cdot h \times h)  \\
               &=& \delta e \cdot \Big( - \sigma R(\omega) + \frac{\Lambda_0}{2} e \times e + D(\omega) h + \frac{\alpha}{2} h \times h \Big)    
                       + \delta \omega \cdot \Big( \frac{1}{\mu} R(\omega) - \sigma T(\omega) + e \times h \Big)       \\
                 && + \delta h \cdot \big( T(\omega) + \alpha e \times h \big)   
                       + d \Big( - \sigma \delta \omega \cdot e + \frac{1}{2\mu} \delta \omega \cdot \omega + \delta e \cdot h \Big)     \\
                 &=& E_{\phi} \delta \phi + d\Theta \,.
\end{eqnarray*}
From the above variation we can simply read equations of motion as follows 
\begin{eqnarray}\label{eom-MMG}
     && T(\omega) + \alpha e \times h = 0 \,, \quad  R(\omega) + \mu e \times h - \sigma \mu T(\omega) = 0 \,,    \nonumber    \\
     && - \sigma R(\omega) + \frac{\Lambda_0}{2} e \times e + D(\omega) h + \frac{\alpha}{2} h \times h = 0 \,.   
\end{eqnarray}
We can also read the symplectic potential $\Theta$ from the variation of the Lagrangian and calculate Noether charge using (\ref{Ncurrent}),
\begin{eqnarray*}
    &&   \Theta = - \sigma \delta \omega \cdot e + \frac{1}{2 \mu} \delta \omega \cdot \omega + \delta e \cdot h   \,,   \\
    &&   Q_{\xi} = - \sigma i_{\xi} \omega \cdot e + \frac{1}{2 \mu} i_{\xi} \omega \cdot \omega + i_{\xi} e \cdot h   \,.     
\end{eqnarray*}
The first equation of (\ref{eom-MMG}) does not guarantee the torsion free condition in this theory. 
So we should make these equations torsion free through shifting connections. 
Shifting connections $\omega$ to new dual spin-connections $\Omega = \omega + \alpha h$, 
then equations of motion (\ref{eom-MMG}) become
\begin{eqnarray}\label{Neom-MMG}
    &&   T(\Omega) = 0 \,,   \nonumber     \\
    &&   R(\Omega) + \frac{\alpha \Lambda_0}{2} e \times e + \mu ( 1+ \sigma \alpha)^2 e \times h = 0   \,,      \\
    &&   D(\Omega) h - \frac{\alpha}{2} h \times h + \sigma \mu (1+\sigma \alpha) e \times h + \frac{\Lambda_0}{2} e \times e = 0  \,.    \nonumber 
\end{eqnarray}
Assuming the frame 1-form fields $e^a$ is invertible and $1+\sigma \alpha \neq 0$ then we can find the field equations condition,
\begin{equation} \label{h-sym}
     e \cdot h = 0  \,,
\end{equation}
i.e. symmetric condition for $h_{\mu\nu}$, using the following identities,
\begin{equation*}
    D(\Omega) T(\Omega) \equiv R(\Omega) \times e  \,,  \quad  D(\Omega) R(\Omega) = 0  \,.
\end{equation*}
Noether charge and symplectic potential can be rearranged using connection shifting $\Omega = \omega + \alpha h$ as follows 
\begin{eqnarray} \label{Q-S-MMG}  
      Q_{\xi} &=& - \sigma i_{\xi} \Omega \cdot e + \frac{1}{2 \mu} i_{\xi} \Omega \cdot \Omega + (1+\sigma \alpha) i_{\xi} e \cdot h   
                    - \frac{\alpha}{2 \mu} ( i_{\xi} \Omega \cdot h + i_{\xi} h \cdot \Omega - \alpha i_{\xi} h \cdot h )   \,,    \nonumber  \\
      \Theta &=& - \sigma \delta \Omega \cdot e + \frac{1}{2 \mu} \delta \Omega \cdot \Omega + (1+\sigma \alpha) \delta e \cdot h    
                    - \frac{\alpha}{2 \mu} ( \delta \Omega \cdot h + \delta h \cdot \Omega - \alpha \delta h \cdot h )    \,.         
\end{eqnarray}
To get the above shifted symplectic potential we used the relation $\delta h \cdot e = \delta e \cdot h$ from the condition (\ref{h-sym}).
%

\subsection{BTZ black hole in MMG}

From now we investigate BTZ black hole which is a solution of MMG theory. The metric of this black hole is the same 
with (\ref{gen-metric}) and frame 1-form fields are also the same with (\ref{dreibein}). 
To find the shifted connection 1-form $\Omega$ we use the first equation of (\ref{eom-MMG}), i.e. $T(\Omega) = de + \Omega \times e = 0$, 
then these connection 1-forms are given by the same form with (\ref{con-1form}).
%
%
Solving the second equation of motion in (\ref{Neom-MMG}) with functions (\ref{BTZ}) we can simply represent the auxiliary fields 
\begin{equation}\label{aux-MMG-2}
   h^a = - \frac{1}{\mu (1 + \sigma \alpha)^2} \Big( \frac{\alpha \Lambda_0}{2} - \frac{1}{2 \ell^2} \Big) e^a = - \lambda e^a  \,,
\end{equation}
where we replace the constant part with $\lambda$ for convenience. Solving third equation of (\ref{Neom-MMG}) we can find a parameter condition
\begin{equation}
       \alpha \lambda^2 + 2 \sigma \mu (1+\sigma\alpha) \lambda - \Lambda_0 = 0 \,.
\end{equation}
Using the above result $h^a = - \lambda e^a$, Noether charges and symplectic potential can be reduced to
\begin{eqnarray}
     Q_{\xi} &=& - \sigma i_{\xi} \Omega \cdot e + \frac{1}{2 \mu} i_{\xi} \Omega \cdot \Omega - \lambda (1+\sigma \alpha) i_{\xi} e \cdot e    
                      + \lambda \frac{\alpha}{2 \mu} ( i_{\xi} \Omega \cdot e + i_{\xi} e \cdot \Omega + \alpha \lambda \, i_{\xi} e \cdot e )   \,,   \nonumber  \\
     \Theta &=& - \sigma \delta \Omega \cdot e + \frac{1}{2 \mu} \delta \Omega \cdot \Omega - \lambda (1+\sigma \alpha) \delta e \cdot e  
                      + \lambda \frac{\alpha}{2 \mu} ( \delta \Omega \cdot e + \delta e \cdot \Omega + \alpha \lambda \, \delta e \cdot e )    \,.            
\end{eqnarray}
Therefore we can represent the charge variation form $\delta \chi_{\xi}$ using above Noether charge and symplectic potential 
\begin{equation}\label{chi-MMG}
     \delta \chi_{\xi} = - \Big( \sigma - \lambda \frac{\alpha}{\mu} \Big) ( i_{\xi} \Omega \cdot \delta e + i_{\xi} e \cdot \delta \Omega )      
                                     + \frac{1}{\mu} i_{\xi} \Omega \cdot \delta \Omega     
                    - \Big( 2\lambda (1+ \sigma \alpha) - \lambda^2 \frac{\alpha^2}{\mu} \Big) i_{\xi} e \cdot \delta e  \,.   
\end{equation}
From now we follow the procedure for BTZ black hole in TMG and NMG.
The non-vanishing variation forms of shifted connection 1-forms related to $d\phi$ component are given by the same results with (\ref{BTZ-var}).
%
%
All variations of the frame fields related to $d\phi$ are vanished. 
Performing the calculation (\ref{chi-MMG}) for the Killing vector $\xi = \frac{\partial}{\partial t}$, the charge variation for BTZ black hole mass in MMG becomes
\begin{equation}
    \delta \chi_{\xi} \Big[ \frac{\partial}{\partial t} \Big] 
              = \Big( \sigma - \lambda \frac{\alpha}{\mu} \Big) \Big( f \delta f + \frac{1}{2} r^3 N \delta N' \Big) d\phi   
        - \frac{1}{\mu} \bigg\{ f \Big( N + \frac{1}{2} r N' \Big) \delta f + \frac{1}{2} r^2 \Big( ff' - \frac{1}{2} r^3 N N' \Big) \delta N' \bigg\} d\phi \,.   
\end{equation}
Applying this result to the definition (\ref{Mass-Ang}), we can obtain the variation of black hole mass 
\begin{eqnarray}\label{M-var-MMG}
    \delta{\cal M} &=& - \frac{1}{8\pi G} \int_{\infty} \delta \chi_{\xi} \Big[\frac{\partial}{\partial t} \Big]    \nonumber  \\
                          &=& \frac{1}{4G} \bigg\{ \frac{1}{\ell^2} \Big( \sigma - \lambda \frac{\alpha}{\mu} \Big) (r_+ \delta r_+ + r_- \delta r_-)     
                                + \frac{1}{\mu \ell^3} (r_- \delta r_+ + r_+ \delta r_-) \bigg\}  \,.
\end{eqnarray}
Substituting the $\lambda$ value (\ref{aux-MMG-2}) into the above formula, then we obtain the mass of BTZ black hole in MMG theory,
\begin{equation}\label{mass-MMG}
   {\cal M} = \frac{r_+^2 + r_-^2}{8G \ell^2} \left( \sigma + \frac{\alpha (1- \alpha \Lambda_0 \ell^2)}{2 \mu^2 \ell^2 (1+ \sigma \alpha)^2} \right)
             + \frac{r_+ r_-}{4G\mu \ell^3}  \,.
\end{equation}
To find the angular momentum of BTZ black hole in MMG theory we consider the charge variation form for the Killing vector 
$\xi = \frac{\partial}{\partial \phi}$. Then the computation of the charge variation form (\ref{chi-MMG}) is given by
\begin{equation}\label{CJ-BTZ-MMG}
    \delta \chi_{\xi} \Big[ \frac{\partial}{\partial \phi} \Big]
        = \bigg\{ \frac{1}{2} \Big( \sigma - \lambda \frac{\alpha}{\mu} \Big) r^3 \delta N'     
          + \frac{1}{\mu} \Big( - f \delta f + \frac{1}{4} r^4 N' \delta N'  \Big) \bigg\} d\phi  \,.
\end{equation}
Adapting the definition of (\ref{Mass-Ang}) the variation form of the angular momentum is given by 
\begin{eqnarray}\label{J-var-MMG}
   \delta {\cal J} &=& \frac{1}{8\pi G} \int_{\infty} \delta \chi_{\xi} \Big[ \frac{\partial}{\partial \phi} \Big]     \nonumber  \\
            &=& \frac{1}{4G} \bigg\{ \frac{1}{\ell} \Big( \sigma - \lambda \frac{\alpha}{\mu} \Big) (r_- \delta r_+ + r_+ \delta r_-)      
                + \frac{1}{\mu \ell^2} (r_+ \delta r_+ + r_- \delta r_-) \bigg\}    \,.
\end{eqnarray}
Substituting the $\lambda$ value (\ref{aux-MMG-2}) into the above formula then we can obtain the angular momentum of BTZ black hole
\begin{equation} \label{ang-MMG}
    {\cal J} = \frac{r_+ r_-}{4G\ell} \left( \sigma + \frac{\alpha (1- \alpha \Lambda_0 \ell^2)}{2 \mu^2 \ell^2 (1+ \sigma \alpha)^2} \right)
                 + \frac{r_+^2 + r_-^2}{8G \mu \ell^2}   \,.
\end{equation}
As already mentioned in case of BTZ black hole in TMG theory, we can re-express the charge variation form (\ref{CJ-BTZ-MMG}) as a total variation form.
Therefore the charge form for the angular momentum of the BTZ black hole in MMG theory can be represented by
\begin{equation}
   \chi_{\xi} \Big[ \frac{\partial}{\partial \phi} \Big]
          = \bigg\{ \frac{1}{2} \Big( \sigma - \lambda \frac{\alpha}{\mu} \Big) r^3 N' + \frac{1}{2\mu} \Big( - f^2 + \frac{1}{4} r^4 N'^2 \Big) \bigg\} \,.
\end{equation}
This result is the same with (\ref{CJ-BTZ-2}) except replacing the coefficient $\sigma$ with $\sigma - \lambda \frac{\alpha}{\mu}$.
Only the constant term contributes to the computation of the angular momentum. So, we can obtain the angular momentum (\ref{ang-MMG}) of BTZ black hole 
in MMG theory. The same results have been calculated by using ADT formalism in \cite{Tekin:2014jna}.

From the metric form of  BTZ black hole (\ref{gen-metric}) and (\ref{BTZ}), we can simply read the angular velocity as
\begin{equation}\label{ang-vel}
    \Omega_{\cal H} = \frac{r_-}{\ell r_+}  \,.
\end{equation}
The entropy of BTZ black hole in MMG have been calculated in \cite{Setare:2015pva}. The Hawking temperature of BTZ black hole is given by 
\begin{equation}\label{H-temp}
     T_{\cal H} = \frac{r_+^2 - r_-^2}{2\pi \ell^2 r_+} \,.
\end{equation}
With (\ref{ang-vel}) and (\ref{H-temp}) the relation between the variation (\ref{M-var-MMG}), (\ref{J-var-MMG}) and 
the variation of the entropy can give the first law of black hole thermodynamics.

\subsection{new type black hole in MMG}
\label{sect:7}

There exists a black hole solution in MMG  at a special point, i.e. ``merger point'' \cite{Arvanitakis:2014yja}. This black hole resembles new type black hole 
solution in NMG theory.
This solution is not locally isometric to the $AdS$ vacuum but asymptotically $AdS$ as $r$ goes to infinity.
There also exists a $dS$ type solution but we pay our attention to $AdS$ type solution. The black hole solution has the same form of (\ref{new-BH}) with (\ref{f-G}).
So we can take same forms of frame 1-form fields $e^a$ (\ref{1f-NMG}), connection 1-forms $\omega^a$ replaced by $\Omega^a$ (\ref{nt-con}).

Solving the second equation of motion in (\ref{Neom-MMG}) with the function (\ref{f-G}), we can determine the auxiliary fields $h^a$ as detailed forms
\begin{eqnarray}\label{aux-newBH} 
   && h^0 = \frac{1}{\mu (1 + \sigma \alpha)^2} \Big( \frac{1}{2\ell^2} - \frac{\alpha \Lambda_0}{2} \Big) e^0  \,,    \nonumber  \\
   && h^1 = \frac{1}{\mu (1 + \sigma \alpha)^2} \Big( \frac{1}{2\ell^2} - \frac{\alpha \Lambda_0}{2} \Big) e^1  \,,        \\
   && h^2 = \frac{1}{\mu (1 + \sigma \alpha)^2} \bigg\{ \frac{1}{2\ell^2} \Big( 1 - \frac{r_+ + r_-}{r} \Big) - \frac{\alpha \Lambda_0}{2} \bigg\} e^2 \,.  \nonumber 
\end{eqnarray}
%
The charge variation (\ref{Q-var}) with Noether charge and symplectic potential (\ref{Q-S-MMG})  can be represented by a simple form
\begin{eqnarray}\label{Q-var-MMG}
    \delta \chi_{\xi} &=& - \sigma (i_{\xi} \Omega \cdot \delta e + i_{\xi} e \cdot \delta \Omega)   
            + \frac{1}{\mu} i_{\xi} \Omega \cdot \delta \Omega + (1 + \sigma \alpha) (i_{\xi} e \cdot \delta h + i_{\xi} h \cdot \delta e)    \nonumber  \\
       && - \frac{\alpha}{\mu} (i_{\xi} \Omega \cdot \delta h + i_{\xi} h \cdot \delta \Omega - \alpha i_{\xi} h \cdot \delta h)  \,.
\end{eqnarray}
To compute this charge variation we need to find non-vanishing variations of frame fields, connection 1-forms and auxiliary fields. 
The $d\phi$ component of all variations of the frame fields vanish. The non-vanishing variations are given by the form
\begin{equation} \label{var-MMG-2}
   \delta \Omega^0 = \delta f d\phi \,, \quad \delta h^2 = \frac{1}{\mu(1 + \sigma \alpha)^2} \delta(ff') d\phi \,.
\end{equation}
%
Because this black hole solution is non-rotating one, so we only need to perform the calculation of the black hole mass. 
To obtain the mass of this black hole we should consider the time-like Killing vector $\xi = \frac{\partial}{\partial t}$. Then the non-vanishing interior products 
of frame 1-forms, connection 1-forms and auxiliary fields with this Killing vector are given by
\begin{equation} \label{int-MMG-n}
      i_{\xi} e^0 = f \,, \quad  i_{\xi} \Omega^2 = ff' \,,  \quad
         i_{\xi} h^0 = \frac{1}{\mu(1 + \sigma\alpha)^2} \bigg\{ \frac{(ff')'}{2} - \frac{\alpha \Lambda_0}{2} \bigg\} f = - \lambda f \,,
\end{equation}
where $\lambda$ is the same constant with (\ref{aux-MMG-2}) whenever we consider the function $f$ of (\ref{f-G}). 
Inserting (\ref{var-MMG-2}) and (\ref{int-MMG-n}) into (\ref{Q-var-MMG}) we can find the charge variation form $\delta \chi_{\xi} [ \frac{\partial}{\partial t} ]$.
But this form of charge variation can be simply reduced to the total variation form, so we can obtain the charge form
\begin{equation} \label{chi-nbh-MMG}
   \chi_{\xi} \Big[ \frac{\partial}{\partial t} \Big]
        = \bigg\{ \frac{1}{2} \Big( \sigma - \lambda \frac{\alpha}{\mu} \Big) f^2 - \frac{\alpha}{ 2\mu^2 (1 + \sigma \alpha)^2} (ff')^2  \bigg\} d\phi \,.
\end{equation}
Computing the third equation of (\ref{Neom-MMG}) provides us with two parameter conditions as follows
\begin{eqnarray} 
    \frac{\alpha}{\mu^2 (1 + \sigma \alpha)^4} \Big( \frac{1}{2\ell^2} - \frac{\alpha \Lambda_0}{2} \Big) 
           - \frac{\sigma}{1 + \sigma \alpha} = 0 \,,   \label{condi-1}    \\
    \frac{\alpha}{\mu^2 (1 + \sigma \alpha)^4} \Big( \frac{1}{2\ell^2} - \frac{\alpha \Lambda_0}{2} \Big)^2 - \frac{\sigma}{\ell^2 (1 + \sigma \alpha)} 
          - \frac{\Lambda_0}{1 + \sigma \alpha} = 0 \,.    \label{condi-2}
\end{eqnarray}   
The above two conditions are nothing but the ``merger point" condition appearing (2.5) in \cite{Arvanitakis:2014yja}.
By the elimination of $\Lambda_0$ these two conditions can be reduced to one condition
\begin{equation}\label{condi-3}
    \frac{\alpha}{\mu^2 \ell^2 (1 + \sigma \alpha)^2} = 2 \sigma + \alpha \,.
\end{equation}
To calculate the charge form (\ref{chi-nbh-MMG}) we need to look over the coefficients of $f^2$ and $(ff')^2$. The parameter $\lambda$ to appear in 
(\ref{chi-nbh-MMG}) is the same value of (\ref{aux-MMG-2}). So we can simplify the coefficient of $f^2$ by using three conditions (\ref{condi-1}), (\ref{condi-2}) 
and (\ref{condi-3}) as a form
\begin{equation}\label{condi-4}
    \sigma - \lambda \frac{\alpha}{\mu} = \sigma + \frac{\alpha (1 - \alpha \Lambda_0 \ell^2)}{2 \mu^2 \ell^2 (1 + \sigma \alpha)^2}  
                   = \sigma + \sigma (1 + \sigma \alpha)      
           = 2\sigma + \alpha = \frac{\alpha}{\mu^2 \ell^2 (1 + \sigma \alpha)^2} \,.
\end{equation}
Therefore the coefficient of $f^2$ in (\ref{chi-nbh-MMG}) are the same that of $(ff')^2$.  So, we can simplify the charge 
\begin{equation}
    \chi_{\xi} \Big[ \frac{\partial}{\partial t} \Big] = 
          \frac{1}{2} \Big( \sigma - \lambda \frac{\alpha}{\mu} \Big) (f^2 - \ell^2 (ff')^2) d\phi   
                  = - \Big( \sigma + \frac{\alpha}{2} \Big) \frac{1}{4\ell^2} (r_+ - r_-)^2 d\phi \,.
\end{equation}
From the above result we can obtain the mass of new type black hole in MMG 
\begin{eqnarray}\label{mass-nbh}
    {\cal M} = - \frac{1}{8\pi G} \int_{\infty} \chi_{\xi} \Big[ \frac{\partial}{\partial t} \Big]            
                 =  \frac{1}{16G\ell^2} \Big( \sigma + \frac{\alpha}{2} \Big) (r_+ - r_-)^2            
                        =  \frac{\alpha}{32 G \mu^2 \ell^4 (1 + \sigma \alpha)^2} (r_+ - r_-)^2 \,.
\end{eqnarray}
This result is described by the square of the difference between two horizons $r_+$ and $r_-$. It is the same form with that of new type black hole in NMG theory 
except the parameter shift $\sigma + \alpha/2$. 
%

\section{Thermodynamic properties of the new type black hole in minimal massive gravity}
In this section we briefly explain to find the central charge of new type black hole and calculate the entropy of this black hole using the method of \cite{Tachikawa:2006sz}. we compare these result with the Cardy formula for black hole entropy. According to the description of AdS/CFT correspondence, we can represent the relation between the black hole mass and the left and right moving energies \cite{Cardy:1986ie,Carlip:1998qw}. This result of black hole mass is the same that we have calculated the above (\ref{mass-nbh}). From these result we can obtain first law of thermodynamics of new type black hole. Then we suggest the relation between the Hawking temperature and the left and right temperatures which come from the left and right moving energies.

\subsection{Boundary central charges}
The method to find the boundary central charge of this MMG theory is described in detail by using the Poisson bracket algebra and 
Hamiltonian analysis \cite{Bergshoeff:2014bia,Bergshoeff:2014pca,Hohm:2012vh,Carlip:2008qh}. The Lagrangian of the three dimensional gravity theories can be represented by 
the Chern-Simons- like form as follows \cite{Bergshoeff:2014bia},
\begin{equation}\label{Lag-CS}
    L_{CSL} = \frac{1}{2} g_{rs} a^r \cdot da^s + \frac{1}{6} f_{rst} a^r \cdot (a^s \times a^t) \,.
\end{equation}
The notation $a^r$ means a collection of Lorentz vector valued 1-forms $a^{ra}_{\mu} dx^{\mu}$, where $r$ is a ``flavor" index running $1 \cdots N$.
Flavor $N$ describes the fields of this MMG theory $e, h$ and $\omega$. $g_{rs}$ is the metric on the flavor space and $f_{rst}$ is the coupling constants. 
This description also includes other gravity theories, i.e. TMG, NMG, ZDG, even MMG theory. Performing the separation of space-time component
\begin{equation}
     a^{ra} = a^{ra}_0 dt + a^{ra}_i  dx^i  \,,
\end{equation}
with writing $\epsilon^{0ij} = \epsilon^{ij}$, the Lagrangian density becomes
\begin{equation}
     {\mathscr L} = - \frac{1}{2} \epsilon^{ij} g_{rs} a^r_i \cdot \dot{a}^s_j + a^r_0 \cdot \phi_r  \,,
\end{equation}
where the Lorentz vectors $a^r_0$ are Lagrange multipliers for primary constraints
\begin{equation}\label{p-const}
     \phi_r = \epsilon^{ij} \Big( g_{rs} \partial_i a^s_{j} + \frac{1}{2} f_{rst} a^s_i \times a^t_j \Big)  \,.
\end{equation}
The Hamiltonian density can be obtained with these primary constraints
\begin{equation}
    {\cal H} = - \frac{1}{2} \epsilon^{ij} g_{rs} a^r_i \cdot \partial_0 a^s_j - {\mathscr L} = - a^r_0 \cdot \phi_r  \,.
\end{equation}
In order to obtain the Poisson brackets of the primary constraints the smeared functionals should be defined by integrating the constraint function (\ref{p-const}) 
against a test function $\xi^r(x)$ as follows
\begin{equation}\label{vphi}
    \varphi[\xi] = \int_{\Sigma} d^2 x \xi^r_a (x) \phi^a_r (x) + Q[\xi] \,,
\end{equation}
where $\Sigma$ is a space-like hypersurface and $Q[\xi]$ is a boundary term to remove delta-function singularities in the brackets of the constraints \cite{Bergshoeff:2014bia}. 
With this  boundary term, $\varphi$ is said to be ``differentiable" under a general variation of the fields. In \cite{Bergshoeff:2014bia} the Poisson brackets of the constraint functions are computed by using
\begin{equation}\label{P-bk}
     \big\{ a^r_{ia} (x) , a^s_{jb} (y) \big\}_{P.B.} = \epsilon_{ij} g^{rs} \eta_{ab} \delta^{(2)} (x-y) \,.
\end{equation}
They are given by 
\begin{equation}
    \big\{ \varphi[\xi] , \varphi[\eta] \big\}_{P.B.} = \varphi[[\xi, \eta]] + \int_{\Sigma} d^2 x \xi^r_a \eta^s_b {\mathscr P}^{ab}_{rs}  
          - \int_{\partial \Sigma} dx^i \xi^r \cdot \Big[ g_{rs} \partial_i \eta^s + f_{rst} (a^s_i \times \eta^t) \Big]   \,,     \label{PB}
\end{equation}
where
\begin{equation}
    [\xi , \eta]^t_c = f^t_{~rs} \epsilon^{ab}_{~~c} \xi^r_a \eta^s_c  \,,
\end{equation}
and
\begin{eqnarray*}
   && {\mathscr P}^{ab}_{rs} = f^t_{~q [r} f_{s] pt} \eta^{ab} \Delta^{pq} + 2 f^t_{~r [s} f_{q] pt} (V^{ab})^{pq} \,,     \\
   && V^{pq}_{ab} = \epsilon^{ij} a^p_{ia} a^q_{jb}  \,, \qquad \Delta^{pq} = \epsilon^{ij} a^p_i \cdot a^q_j  \,.
\end{eqnarray*}
The Poisson brackets of the first-class constraints with the dynamical fields of the theory generate local Lorentz transformations and diffeomorphisms. The second term of (\ref{PB}) is related to the secondary constraint associated with the consistency condition guaranteeing time independence of the primary constraints. If we regard the test function $\xi^r_a (x)$ as the gauge parameters of boundary condition preserving gauge transformation, then the Poisson bracket algebra is isomorphic to the Lie algebra of the asymptotic symmetries and generally can be centrally extended \cite{Brown:1986nw}. 
%

%
The Chern-Simons-like model (\ref{Lag-CS}) is manifestly invariant under diffeomorphism and local Lorentz transformations. 
To understand these constraints which generate these symmetries, it is convenient to investigate the Poisson brackets of the gauge 
transformation with the dynamical variables of the theory. The calculations can be performed using (\ref{p-const})(\ref{vphi}) and (\ref{P-bk}) 
as a form
\begin{equation}\label{vari-Po}
    \{ \varphi [\xi] \,, a^r_i \} = \partial_i \xi^r + f^r_{~st} a^s_i \times \xi^t \,.
\end{equation}
The constraint $\varphi_{\omega}[\xi^{\omega}]$ generate a local Lorentz transformation with $\xi^{\omega} = \zeta$ and $\xi^r = 0$ for $r \neq \omega$, as follows
\begin{eqnarray}
      && \delta_{\zeta} \omega_i = \{ \varphi_{\omega} [\zeta] \,, \omega_i \} = \partial_i \zeta + \omega_i \times \zeta \,,   \label{vari-a}   \\
      && \delta_{\zeta} e_i = \{ \varphi_{\omega} [\zeta] \,, e_i \} = a^r_i \times \zeta \,.      \label{vari-b}
\end{eqnarray}
In (\ref{vari-b}), we have used a relation $f^r_{~s\omega} = g^{r t} f_{t s \omega} = \delta^r_{~s}$ where a coupling constant
$f_{r s\omega}$ is given by $g_{r s}$. Diffeomorphisms associated with a vector field $\zeta^{\mu}$ are generated by an appropriate combinations of constraint
functions
\begin{equation}
    \varphi_{\rm diff} [\zeta] = \sum_r \varphi_r[\zeta^{\mu} a^r_{\mu}] \,.
\end{equation}
Then, by using (\ref{vari-Po}) we can find 
\begin{equation}\label{vari-diff}
    \delta_{\zeta} a^r_i = \{ \varphi_{\rm diff} [\zeta] \,, a^r_i \} = \pounds_{\zeta} a^r_i + \cdots \,,
\end{equation}
where $\pounds_{\zeta}$ describes the Lie derivative with respect to the vector field $\zeta^{\mu}$ and $\cdots$ means the term proportional to equations 
of motion. So, on-shell this gives the correct transformation rule for the dynamical variables of the theory.

%
%


%
To find the boundary central charges from the Poisson bracket algebra (\ref{PB}) we need to consider the two sets of mutually commuting first class constraints 
\begin{equation}
      L_{\pm} [\zeta] = \varphi_e [\zeta^{\mu} e_{\mu}] + \varphi_h [\zeta^{\mu} h_{\mu}] + a_{\pm} \varphi_{\omega} [\zeta^{\mu} e_{\mu}]  \,,
\end{equation}
with constant $a_{\pm}$. Then the Poisson brackets should be
\begin{eqnarray}
      && \big\{  L_{\pm} [\xi] , L_{\pm} [\eta] \big\} = \mp \frac{2}{\ell} L_{\pm} [[\xi, \eta]]    \,,     \label{L1}    \\
      && \big\{ L_+[\xi] , L_- [\eta] \big\}  = 0  \,.     \label{L2}
\end{eqnarray}
To calculate above commutators, we firstly need to determine the non-zero components of the metric $g_{r s}$ and the coupling constants $f_{r s t}$ on the "flavor" space in MMG theory. They are given by
\begin{eqnarray}
    && g_{e \omega} = - \sigma \,, ~~~ g_{h e} = 1 \,, ~~~ g_{\omega \omega} = \frac{1}{\mu}  \,,    \nonumber    \\
    && f_{e \omega \omega} = - \sigma \,, ~~~  f_{e e e} = \Lambda_0 \,, ~~~ f_{h \omega e} = 1 \,,     \\
    && f_{\omega \omega \omega} = \frac{1}{\mu} \,, ~~~ f_{e h h} = \alpha \,.      \nonumber 
\end{eqnarray}
Using (\ref{PB}), the Poisson brackets of the primary constraints can be determined as follows,
\begin{eqnarray}
    && \{ \phi^a_{\omega} \,, \phi^b_{\omega} \}_{P.B.} = \varepsilon^{ab}_{~~c} \phi^c_{\omega} \,,  ~~~ 
              \{ \phi^a_{\omega} \,, \phi^b_e \}_{P.B.} = \varepsilon^{ab}_{~~c} \phi^c_e \,,      \nonumber   \\
    && \{ \phi^a_{\omega} \,, \phi^b_h \}_{P.B.} = \varepsilon^{ab}_{~~c} \phi^c_h \,,   ~~~
             \{ \phi^a_e \,, \phi^b_e \}_{P.B.} = \Lambda_0 \varepsilon^{ab}_{~~c} \phi^c_h \,,    \nonumber  \\
    && \{ \phi^a_h \,, \phi^b_h \}_{P.B.} = \alpha \varepsilon^{ab}_{~~c} \phi^c_{\omega} \,,     \nonumber   \\
    && \{ \phi^a_e \,, \phi^b_h \}_{P.B.} = \alpha \varepsilon^{ab}_{~~c} \phi^c_e   
         + \mu(1+\sigma\alpha) \varepsilon^{ab}_{~~c} \phi^c_{\omega} + \sigma \mu (1+\sigma\alpha) \varepsilon^{ab}_{~~c} \phi^c_h \,.
\end{eqnarray}
Let $\bar{e}$ be the $AdS$ background dreibein. By using the ``merger point" conditions (\ref{condi-1}), (\ref{condi-2}), the auxiliary field (\ref{aux-newBH}) becomes $\bar{h} = \beta \bar{e}$ where the parameter $\beta$ is given by
\begin{equation}
   \beta = \frac{\sigma \mu (1 + \sigma \alpha)}{\alpha} \,,
\end{equation}
on the $AdS$ background. Also,  for the new type black hole case, we can find the Poisson brackets between smeared functionals
\begin{eqnarray}
   && \{ \varphi_e [\xi^e] \,, \varphi_e [\eta^e] \}_{P.B.} = \Lambda_0 \varphi_h [[ \xi \,, \eta]] \,,  ~~   
         \{ \varphi_h [\xi^h] \,, \varphi_h [\eta^h] \}_{P.B.} = \beta^2 \alpha \varphi_h [[ \xi \,, \eta]] \,, ~~ \nonumber   \\
   && \{ \varphi_h [\xi^h] \,, \varphi_{\omega} [\eta^e] \}_{P.B.} = \beta \varphi_h [[ \xi \,, \eta]] \,,  ~~ 
         \{ \varphi_{\omega} [\xi^e] \,, \varphi_e [\eta^e] \}_{P.B.} = \varphi_e [[ \xi \,, \eta]] \,,  ~~      \\
   && \{ \varphi_{\omega} [\xi^e] \,, \varphi_{\omega} [\eta^e] \}_{P.B.} = \varphi_{\omega} [[ \xi \,, \eta]] \,,  ~~   \nonumber   \\
   && \{ \varphi_e [\xi^e] \,, \varphi_h [\eta^h] \}_{P.B.} =  \beta \mu (1+\sigma\alpha) \varphi_{\omega} [[\xi \,, \eta]]   
              + \beta \alpha \varphi_e[[\xi \,, \eta]] + \beta \sigma \mu (1+\sigma\alpha) \varphi_h [[\xi \,, \eta]]   \,,     \nonumber  
\end{eqnarray}
where $\xi^h = \xi^{\mu} h_{\mu} = \beta \xi^e$ and $[\xi \,, \eta] = \xi^e \times \eta^e$. With the parameter
\begin{equation}
     \Lambda_0 = - \frac{\mu^2 (1+\sigma\alpha)^2}{\alpha} = - \alpha \beta^2 \,,
\end{equation}
calculated from (\ref{condi-1}) and (\ref{condi-2}), we can take advantage of (\ref{L2}) 
\begin{eqnarray}
     \{ L_+[\xi] \,, L_-[\eta] \} &=&  (2\beta\mu (1+\sigma \alpha) + a_+ a_-) \varphi_h [[\xi \,, \eta]]     \nonumber   \\
       && + (2\beta\alpha + a_+ + a_-) \big( \varphi_e [[\xi \,, \eta]] + \beta \varphi_h [[\xi \,, \eta]] \big) = 0 \,,    
\end{eqnarray}
to find conditions for two parameters $a_+$ and $a_-$. This commutator gives two equations
\begin{eqnarray}
    && 2 \beta \alpha + a_+ + a_- = 0\,,    \\  
    && 2\beta\mu(1+\sigma\alpha) + a_+ a_- = 0 \,.
\end{eqnarray}
So, we can obtain constants 
\begin{equation}
    a_{\pm} = -\beta \alpha \left( 1 \pm \sqrt{1+\frac{2\mu(1+\sigma\alpha)}{\beta\alpha^2}} \right)   
                  = - \sigma \mu (1 + \sigma \alpha) \left( 1 \pm \sqrt{1 + \frac{2\sigma}{\alpha}} \right)  \,.
\end{equation}
Therefore we can use the above parametrization for $a_{\pm}$ and two identities 
\begin{eqnarray}
     && \!\! a_{\pm} = -\sigma\mu(1+\sigma\alpha) \pm \frac{1}{\ell} = -\beta\alpha \pm \frac{1}{\ell} \,, \\
     && \!\! a_{\pm}^2 + 2\beta\mu(1+\sigma\alpha) = 2\frac{\sigma\mu^2 (1+\sigma\alpha)^2}{\alpha} = \pm \frac{2}{\ell} a_{\pm} \,,
\end{eqnarray}
to find out the Poisson bracket
\begin{eqnarray}
    \big\{  L_{\pm} [\xi] , L_{\pm} [\eta] \big\} &=& \mp \frac{2}{\ell} L_{\pm} [[\xi, \eta]]       \\
          && \mp \frac{2}{\ell} \Big( 2\sigma + \alpha \pm \frac{1}{\mu\ell} \Big) \int_{\partial \Sigma} d\phi \xi^e \cdot 
                         \bigg[ \partial_{\phi} \eta^e + \bigg( \bar{\omega}_{\phi} 
                         \pm \Big( \frac{2}{\ell} \mp a_{\pm} \Big) \bar{e}_{\phi} \bigg) \times \eta^e \bigg]  \,.    \nonumber 
\end{eqnarray}
After including the proper normalizations of first class constraints, we can find the asymptotic symmetry algebra consisting of two sets of Virasoro algebra 
with a central charge
\begin{equation}\label{cent-nbh}
     c_{L, R} = \frac{3\ell}{2G} \Big( 2\sigma + \alpha \pm \frac{1}{\mu\ell} \Big)  \,,
\end{equation}
where $G$ is the three dimensional Newton constant. We notice that the central charges can not be reduced to the TMG representations in the TMG limit 
$\alpha \rightarrow 0$, because there exist a relation (\ref{condi-3}) between $\mu$ and $\alpha$ at ``merger point"  in MMG theory. 
This is somewhat different case from the $AdS$ background one since the solution of  new type black hole in TMG does not exist. 
The same result can also be obtained by using the variation of the boundary charge $Q_{\pm} [\xi]$ 
\begin{eqnarray}
   \delta Q_{\pm} [\xi] &=& - \int_{\partial \Sigma} dx^i \Big( g_{es} \xi^{e} + g_{hs} \xi^{h} +  a_{\pm} g_{\omega s} \xi^{e} \Big) \cdot \delta a_i^s   \nonumber \\
         &=& \Big( 2\sigma + \alpha \pm \frac{1}{\mu\ell} \Big) \int_{\partial \Sigma} d\phi \xi^e 
                       \cdot \bigg( \delta \bar{\omega}_{\phi} \pm \Big( \frac{2}{\ell} \mp a_{\pm} \Big)  \delta \bar{e}_{\phi} \bigg)     \,, 
\end{eqnarray}
where $s$ sums over $e$, $h$ and $\omega$. With some proper constants the coefficient of the integral give the same central charges.

\subsection{Thermodynamics}

To find the thermodynamic relations for new type black hole in MMG theory, we should use the previous result for the mass in section \ref{sect:7} and 
calculate the entropy of the black hole. To obtain the black hole entropy of new type black hole, we firstly consider the Lagrangian for 
the Chern-Simons-like form (\ref{Lag-CS}).  The variation for this Lagrangian is given by
\begin{equation}
    \delta L_{CSL} = d \Big( \frac{1}{2} g_{rs} \delta a^r \cdot a^s \Big) + \delta a^r \cdot \Big( g_{rs} da^s + \frac{1}{2} f_{rst} a^s \times a^t \Big) \,,
\end{equation}
where the first term gives the symplectic potential 
\begin{equation}
    \Theta (a\,, \delta a) = \frac{1}{2} g_{rs} \delta a^r \cdot a^s  \,,
\end{equation}
and the second term describes equations of motion.
Following the Wald's formalism, we can find Noether charge
\begin{equation}
    Q_{\xi} = \frac{1}{2} g_{rs} i_{\xi} a^r \cdot a^s  \,,
\end{equation}
with the on-shell condition. Let us consider the variation of Noether charge
\begin{equation}\label{Q-vari}
     \delta Q_{\xi} = \frac{1}{2} \big( g_{rs} i_{\xi} \delta a^r \cdot a^s + g_{rs} \delta a^r \cdot i_{\xi} a^s \big) \,.
\end{equation}
The Killing vector $\xi$ vanishes on the bifurcation horizon $\mathcal H$, so the interior product of the symplectic potential should also be vanished, i.e.,
$i_{\xi} \Theta = 0$. By using this condition, the first term of the variation form (\ref{Q-vari}) becomes the same one with the second term. So, we can get 
\begin{equation}
    \delta Q'_{\xi} = g_{rs} i_{\xi} a^r \cdot \delta a^s  \,,
\end{equation}
which is the same form of the charge variation form (\ref{Q-var}) on the bifurcation horizon $\mathcal H$, i.e., $\delta Q'_{\xi} = \delta \chi_{\xi}$. 
Therefore, we can define the variation form of the entropy of a black hole in the Chern-Simons-like gravity theory
\begin{equation}\label{def-Ent}
    \frac{\kappa}{2 \pi} \delta {\mathcal S}_{BH} = - \frac{1}{8\pi G} \int_{\mathcal H} \delta \chi_{\xi} \,.
\end{equation}
%
%
The calculation for the black hole entropy including the gravitational Chern-Simons term in the action has been performed in \cite{Tachikawa:2006sz}. 
The charge for the black hole entropy has been defined by $Q'_{\xi} = Q_{\xi} - C_{\xi}$ where $\delta C_{\xi} = i_{\xi} \Theta + \Sigma_{\xi}$ 
with a choice $\Sigma_{\xi}=0$. So, it is the same definition with (\ref{def-Ent}). Now we look for the entropy of new type black hole in MMG theory. 
Then the entropy formula can be written by the variational form
\begin{equation}\label{v-ent}
   \delta {\cal S}_{BH} = - \frac{1}{8\pi G} \cdot \frac{2\pi}{\kappa} \int_{\cal H} \left( \delta \chi_{\xi} \Big[ \frac{\partial}{\partial t} \Big]
                + \Omega_{\cal H} \delta \chi_{\xi} \Big[ \frac{\partial}{\partial \phi} \Big] \right)  \,,
\end{equation}
which calculation should be performed at the event horizon of a black hole ${\cal H}$ and $\Omega_{\cal H}$ is the angular velocity 
at the horizon when we consider a rotating black hole case. Therefore we just consider the first variational charge form. 
By using (\ref{chi-nbh-MMG}), (\ref{condi-3}) and (\ref{condi-4}), we can obtain the charge variation form 
\begin{equation}\label{vchi-nbh}
     \delta \chi_{\xi} \Big[ \frac{\partial}{\partial t} \Big] = - \frac{1}{2\ell^2} \Big( \sigma + \frac{\alpha}{2} \Big) (r_+ - r_-) (\delta r_+ - \delta r_-) d\phi  
           = - \frac{\alpha (r_+ - r_-)}{4\mu^2 \ell^4 (1+\sigma \alpha)^2} (\delta r_+ - \delta r_-) d\phi \,,
\end{equation}
at the event horizon $r=r_+$. The Hawking temperature is given by 
\begin{equation}\label{HT}
     T_H = \frac{\kappa}{2\pi} = \frac{r_+ - r_-}{4\pi \ell^2} \,.
\end{equation}
Substituting (\ref{vchi-nbh}) and (\ref{HT}) into (\ref{v-ent}) and then integrate the variational form, we can obtain the entropy of new type black hole in MMG theory
as follows
\begin{equation}\label{ent-nbh}
     {\cal S}_{BH} = \frac{\pi}{2G} \Big( \sigma + \frac{\alpha}{2} \Big) (r_+ - r_-) 
                    = \frac{\pi \alpha (r_+ - r_-)}{4G \mu^2 \ell^2 (1+\sigma \alpha)^2} \,.
\end{equation}
Considering the mass formula (\ref{mass-nbh}), the only physical parameter is the mass ${\cal M}$ of the new type black hole and can also be given by 
the difference between two horizons $r_+ - r_-$. So we can obtain the first law of black hole thermodynamics by using the variation of mass (\ref{mass-nbh}) and (\ref{ent-nbh}) 
\begin{equation}\label{1st-nbh}
      \delta {\cal M} = T_H \delta {\cal S}_{BH}  \,.
\end{equation}
According to the usual description of AdS/CFT correspondence, the black hole entropy can be represented by the Cardy formula 
which depicts the entropy of a dual CFT in terms of central charges $c_L$ and $c_R$ at the temperatures $T_L$ and $T_R$ as follows
\begin{equation}\label{cardy}
    {\cal S}_{BH} = \frac{\pi^2 \ell}{3} \big( c_L T_L + c_R T_R \big)  \,.
\end{equation}
In three dimensional gravity theories, the mass of an asymptotically $AdS$ black hole can be obtained by
\begin{equation}\label{M-E}
     {\cal M} = {\cal E}_L + {\cal E}_R \,, 
\end{equation}
where the left and right moving energies of the dual CFT can be defined as 
\begin{equation}
      {\cal E}_L = \frac{\pi^2 \ell}{6} c_L T_L^2 \,,  \quad  {\cal E}_R = \frac{\pi^2 \ell}{6} c_R T_R^2  \,.
\end{equation}
Here, we suggest that the left moving and right moving temperatures should be equal to the Hawking temperature such as
\begin{equation}\label{T-eq}
    T_L = T_R = T_H \,,
\end{equation}
then relations (\ref{cardy}) and (\ref{M-E}) are given by the same results (\ref{ent-nbh}) and (\ref{mass-nbh}) with the central charges (\ref{cent-nbh}).
Therefore the Smarr formula is given by 
\begin{equation}\label{smarr-nbh}
     {\cal M} = \frac{1}{2} T_H {\cal S}_{BH} \,.
\end{equation}
From the variation of this formula with $r_+ - r_-$ we can also derive the first law of black hole thermodynamics (\ref{1st-nbh}).

\section{Conclusion}

In this paper we have investigated the variation form (\ref{Mass-Ang}) of the mass and angular momentum of black holes through the Wald formalism. 
These variations are satisfied with the first law of black hole thermodynamics (\ref{1st-law}). It have been shown that the Lagrangian for diverse three dimensional gravity theories can be expressed by using first order formalism which comprise with auxiliary fields \cite{Bergshoeff:2014bia,Bergshoeff:2014pca,Afshar:2014ffa}. 
Using the definition of these variations (\ref{Mass-Ang}), 
we have obtained masses and angular momenta of the asymptotically $AdS_3$ black holes in three dimensional gravity theories with the first order formalism.
Following the Wald formalism, we have calculated symplectic potential and conserved Noether charge which consist of dreibeins, spin connections and auxiliary fields. To find conserved charges such as mass and angular momentum, we need to calculate the value (\ref{Q-var}). Firstly, we have reproduced well-known results of the mass and angular momentum for some black holes in TMG and NMG as some examples. Secondly we have paid our attention to compute new results of mass and angular momentum of BTZ and new type black holes in MMG theory. 
These results are the same form with one in TMG theory with a modified $\sigma$ parameter which is a function of $\alpha$ and $\Lambda_0$. 
The mass of new type black hole in MMG theory is also the same form 
with that of NMG theory. Only the difference is a parameter $\sigma+ \alpha/2$ which include a parameter $\alpha$ related to the auxiliary field $h$ term in the action.

The MMG theory is an alternative modified theory to avoid ``bulk vs. boundary clash'' \cite{Bergshoeff:2014pca}. 
MMG equation (\ref{MMG-eq}) can be expressed by a simple modification of equation in TMG theory including an auxiliary field term with a parameter $\alpha$. The elimination of the 
auxiliary fields from the MMG action cannot make an action for the metric only. So in order to get proper conserved charges such as mass and angular momentum, we should make a certain formula including auxiliary fields even though some results have obtained by using the equation (\ref{MMG-eq}) in 
\cite{Setare:2015vea}. Definitions (\ref{Mass-Ang}) are enough to find mass and angular momentum in three dimensional gravity theories while we are dealing with the integration at spatial infinity. 

We also calculate the central charges of new type black hole by investigation of the Poisson brackets of the constraint functions coming from the ``Chern-Simons-like" Lagrangian form. We compute the entropy of new type black hole by taking use of the method in \cite{Tachikawa:2006sz} with the first-order formalism. From these results, we can read the Smarr relation (\ref{smarr-nbh}) which is also derived from the first law of black hole thermodynamics (\ref{1st-nbh}). The entropy and mass can be compare with the Cardy formula ({\ref{cardy}}) and energy definition of the dual CFT (\ref{M-E}). This comparison leads us to suggest a relation (\ref{T-eq}) that the left and right moving temperatures should be equal to the Hawking temperature. Then the Smarr relation is satisfied under this suggestion.

Many constructions of the conserved charge for black holes are introduced in a review paper \cite{Adami:2017phg}. By using the linearization of the metric on the asymptotically 
flat spacetime, the conserved charge has been constructed by Arnowitt, Deser and Misner (ADM) \cite{Arnowitt:1962hi}. This ADM formalism has been extended to the covariant and 
higher curvature gravity theories by Abbott, Deser and Tekin \cite{Abbott:1981ff,Deser:2002rt,Deser:2002jk}. There was a non-trivial generalization of the ADT charge formalism by promoting to the off-shell level and 
including non-covariant term like a gravitational Chern-Simons term \cite{Kim:2013zha,Kim:2013cor}. In \cite{Adami:2017phg} the quasi-local conserved charge 
for the Chern-Simons-like theories of gravity (\ref{Lag-CS}) has been constructed with the first order formalism. 
To find this charge the authors of \cite{Adami:2017phg} have used the off-shell ADT method and the field variation with the Lorentz-Lie derivative 
\begin{equation}
     \delta_{\xi} a^{ra} =  {\mathfrak L}_{\xi} a^{ra} - \delta^r_{~\omega} d \tilde{\lambda}^a \,,
\end{equation}
where the definition of the Lorentz-Lie derivative ${\mathfrak L}$ is given by 
\begin{equation}
     {\mathfrak L}_{\xi} e^a = \pounds_{\xi} e^a + \lambda^a_{~c} e^c \,,
\end{equation}
and $\tilde{\lambda}^a = 1/2 \cdot \epsilon^a_{~bc} \lambda^{bc}$ is a generator of the local Lorentz transformation.
This derivative has been introduced to avoid the divergence of the spin connection on the event horizon,
even though the interior product between spin connection and Killing vector becomes finite on the bifurcation surface \cite{Jacobson:2015uqa}.

In our approach, the Wald formalism for the diffeomorphism invariant Lagrangian has been adapted. Because the variation of the field variables 
for a vector filed $\xi$ can be described by (\ref{vari-diff}), the symplectic current $\omega(\phi, \delta \phi, \pounds_{\xi} \phi)$ 
in (\ref{var-H}) vanishes when we take $\pounds_{\xi} \phi = 0$. So, we define the charge variational form as follow the path of the solution
on the Cauchy surface $\Sigma$.


\section*{Acknowledgement}
S.N was supported by Basic Science Research Program through the National Research Foundation of Korea (NRF) funded by the Ministry of Education 
(No.2013R1A1A2004538). J.D.P was supported by a grant from the Kyung Hee University in 2009 (KHU-20110060) and Basic Science Research Program 
through the National Research Foundation of Korea (NRF) funded by the Ministry of Education (No.2015R1D1A1A01061177).


\appendix 
\section*{\Large \bf Appendices}
 
\renewcommand{\theequation}{A.\arabic{equation}}
\setcounter{equation}{0}

\section{Some calculations for the curvature two-forms of the warped $AdS_3$ black hole}
%
 In this appendix we summarize some useful functions and relations to calculate curvature 2-forms of the warped $AdS_3$ black hole.  
This black hole solution is represented by some functions given by (\ref{wap-func}). 
The second function of (\ref{wap-func}) can be represented by the form
 \begin{equation}
       RN= \frac{\ell \sqrt{(\nu^2 + 3)}}{2} \sqrt{(r - r_+)(r - r_-)}  \,.
 \end{equation}
With all functions of (\ref{wap-func}) and their derivatives we can calculate curvature 2-forms.
From now we abbreviate all functions as their abridged form without coordinate $r$ including the above formulae. 
Substituting connections (\ref{con-wap}) into the second equation in (\ref{T-R}), curvature two-forms are given by
\begin{eqnarray} \label{cur-warp}
      R^0 &=& e^1 \wedge e^2  \bigg[ \Big( \frac{R^2 N^{\theta \prime}}{\ell} \Big)^2 
                      + \frac{2(NR')'}{\ell^2} \cdot \frac{2RN}{\ell^2} + \Big( \frac{2NR'}{\ell^2} \Big)^2 \bigg]       \nonumber  \\
                && + e^1 \wedge e^0  \bigg[ \frac{(R^2 N^{\theta \prime})'}{\ell} \cdot \frac{2RN}{\ell^2} 
                      + \frac{2NR'}{\ell^2} \cdot \frac{2 R^2 N^{\theta \prime}}{\ell} \bigg]  \,,   \nonumber    \\
      R^1 &=& e^2 \wedge e^0 \bigg[ - \Big( \frac{R^2 N^{\theta \prime}}{\ell} \Big)^2 
                      - \frac{2RN'}{\ell^2} \cdot \frac{2NR'}{\ell^2} \bigg]     \,,                \\
      R^2 &=& e^0 \wedge e^1 \bigg[ 3 \Big( \frac{R^2 N^{\theta \prime}}{\ell} \Big)^2 - \frac{2(RN')'}{\ell^2} \cdot \frac{2RN}{\ell^2}
                      - \Big( \frac{2RN'}{\ell^2} \Big)^2 \bigg]      \nonumber       \\
                && + e^1 \wedge e^2 \bigg[ - \frac{(R^2 N^{\theta \prime})'}{\ell} \cdot \frac{2RN}{\ell^2} 
                      - \frac{2 R^2 N^{\theta \prime}}{\ell} \cdot \frac{2NR'}{\ell^2} \bigg]    \,.    \nonumber   
 \end{eqnarray}
Solving the second equation of (\ref{eoms}) with torsion free condition, then we obtain the auxiliary fields as follows
\begin{eqnarray}
    h^0 &=& \frac{1}{\mu} \bigg\{ - \frac{3}{2} \Big( \frac{R^2 N^{\theta \prime}}{\ell} \Big)^2
                         - \frac{1}{2} \bigg( \frac{2(N R')'}{\ell^2} \cdot \frac{2RN}{\ell^2}      
                        + \Big( \frac{2N R'}{\ell^2} \Big)^2 \bigg)  
                        + \frac{1}{2} \cdot  \frac{2R N'}{\ell^2} \cdot \frac{2N R'}{\ell^2}     \nonumber  \\
                 &&   + \frac{1}{2} \bigg( \frac{2(R N')'}{\ell^2} \cdot \frac{2RN}{\ell^2} + \Big( \frac{2R N'}{\ell^2} \Big)^2  \bigg) \bigg\} e^0   
                        - \frac{1}{\mu} \bigg\{ \frac{(R^2 N^{\theta \prime})'}{\ell} \cdot \frac{2RN}{\ell^2} 
                         + \frac{2N R'}{\ell^2} \cdot \frac{2R^2 N^{\theta \prime}}{\ell} \bigg\} e^2  \,,     \nonumber   \\
%
    h^1 &=& \frac{1}{\mu} \bigg\{ - \frac{3}{2}\Big( \frac{R^2 N^{\theta \prime}}{\ell} \Big)^2 
                         + \frac{1}{2} \bigg( \frac{2(N R')'}{\ell^2} \cdot \frac{2RN}{\ell^2}      
                         + \Big( \frac{2N R'}{\ell^2} \Big)^2 \bigg) - \frac{1}{2} \frac{2R N'}{\ell^2} \cdot \frac{2N R'}{\ell^2}     \nonumber        \\      
                &&   + \frac{1}{2} \bigg( \frac{2(R N')'}{\ell^2} \cdot \frac{2RN}{\ell^2} + \Big( \frac{2R N'}{\ell^2} \Big)^2  \bigg) \bigg\} e^1 \,,    \label{aux-warp}    \\
%
    h^2 &=& \frac{1}{\mu} \bigg\{ \frac{5}{2} \Big( \frac{R^2 N^{\theta \prime}}{\ell} \Big)^2 
                         + \frac{1}{2} \bigg( \frac{2(N R')'}{\ell^2} \cdot \frac{2RN}{\ell^2}                 
                         + \Big( \frac{2N R'}{\ell^2} \Big)^2 \bigg) + \frac{1}{2} \cdot  \frac{2R N'}{\ell^2} \cdot \frac{2N R'}{\ell^2}    \nonumber   \\
                &&    - \frac{1}{2} \bigg( \frac{2(R N')'}{\ell^2} \cdot \frac{2RN}{\ell^2} + \Big( \frac{2R N'}{\ell^2} \Big)^2  \bigg) \bigg\} e^2       
                         + \frac{1}{\mu} \bigg\{ \frac{(R^2 N^{\theta \prime})'}{\ell} \cdot \frac{2RN}{\ell^2} 
                         + \frac{2N R'}{\ell^2} \cdot \frac{2R^2 N^{\theta \prime}}{\ell} \bigg\} e^0  \,.    \nonumber       
\end{eqnarray}
We firstly consider the differentiation of all functions of (\ref{wap-func}) with respect to $r$. The derivatives of these functions are given by
\begin{eqnarray}
   R' &=& \frac{R}{2r} + \frac{3(\nu^2 - 1)}{8} \frac{r}{R}   \,,      \nonumber   \\
   N' &=& \frac{\ell \sqrt{(\nu^2 + 3)}}{4} \bigg\{ \frac{1}{R} \frac{r^2 - r_+ r_-}{r \sqrt{(r - r_+)(r - r_-)}} 
                - \frac{3(\nu^2 - 1)}{4} \frac{r \sqrt{(r - r_+)(r - r_-)}}{R^3} \bigg\} \,,            \\
   N^{\theta \prime} &=& \frac{\sqrt{r_+ r_- (\nu^2 + 3)}}{2} \frac{1}{r R^2}        
               - \frac{3(\nu^2 -1)}{8} \frac{r}{R^4} \big(2\nu r - \sqrt{r_+ r_- (\nu^2 + 3)} \big)     \,.     \nonumber 
\end{eqnarray}
From the above formulae and (\ref{wap-func}) we can get some relations
\begin{eqnarray}\label{Funcs}    
     \frac{R^2 N^{\theta \prime}}{\ell} &=& \frac{\sqrt{r_+ r_- (\nu^2 + 3)}}{2 \ell} \frac{1}{r}   
                       - \frac{3(\nu^2 -1)}{8 \ell} \frac{r}{R^2} \big(2\nu r - \sqrt{r_+ r_- (\nu^2 + 3)} \big)  \,,    \nonumber   \\
%
     \frac{2NR'}{\ell^2} &=& \frac{\sqrt{(\nu^2 + 3)}}{2 \ell} \sqrt{(r - r_+)(r - r_-)}   
                    \cdot \Big( \frac{1}{r} + \frac{3(\nu^2 - 1)}{4} \frac{r}{R^2} \Big)    \,,      \\
%
     \frac{2RN'}{\ell^2} &=& \frac{\sqrt{(\nu^2 + 3)}}{2 \ell} \bigg\{ \frac{2r - r_+ - r_-}{\sqrt{(r - r_+)(r - r_-)}}   
                      - \sqrt{(r - r_+)(r - r_-)} \Big( \frac{1}{r} + \frac{3(\nu^2 - 1)}{4} \frac{r}{R^2} \Big) \bigg\} \,.      \nonumber    
\end{eqnarray}
and their derivatives as follows
\begin{eqnarray}\label{Derivs}
      \Big(\frac{R^2 N^{\theta \prime}}{\ell} \Big)' &=& - \frac{\sqrt{r_+ r_- (\nu^2 + 3)}}{2 \ell} \frac{1}{r^2}  
                    - \frac{3(\nu^2 - 1)}{8 \ell} \bigg[ \frac{2\nu r}{R^2}          
                    - \frac{3(\nu^2 - 1)}{4} \frac{r^2}{R^4} (2\nu r - \sqrt{r_+ r_- (\nu^2 + 3)}) \bigg]   \,,     \nonumber    \\
%
      \Big( \frac{2NR' }{\ell^2}\Big)' &=& \frac{\sqrt{(\nu^2 + 3)}}{2 \ell} \bigg[ \frac{(r_+ + r_-)r - 2 r_+ r_-}{2r^2 \sqrt{(r - r_+)(r - r_-)}}   
                  + \frac{3(\nu^2 - 1)}{4} \bigg\{ \frac{r(2r - r_+ - r_-)}{2R^2 \sqrt{(r - r_+)(r - r_-)}}          \nonumber   \\ 
                  && - \frac{3(\nu^2 - 1)}{4} \frac{r^2}{R^4} \sqrt{(r - r_+)(r - r_-)} \bigg\} \bigg]  \,,              \\
%
      \Big( \frac{2RN'}{\ell^2} \Big)' &=& \frac{\sqrt{(\nu^2 + 3)}}{2 \ell} \bigg[ \frac{r^2 + r_+ r_-}{r^2 \sqrt{(r - r_+)(r - r_-)}}    
                        - \frac{(r^2 - r_+ r_-)(2r - r_+ - r_-)}{2r [(r - r_+)(r - r_-)]^{3/2}}     \nonumber   \\
                  && - \frac{3(\nu^2 - 1)}{4} \bigg\{ \frac{r(2r - r_+ - r_-)}{2R^2 \sqrt{(r - r_+)(r - r_-)}}    
                        - \frac{3(\nu^2 - 1)}{4} \frac{r^2}{R^4} \sqrt{(r - r_+)(r - r_-)} \bigg\} \bigg]  \,.      \nonumber     
\end{eqnarray}
Now we use the previous results to calculate curvature 2-forms (\ref{cur-warp}). 
The calculations of the right hand side of (\ref{cur-warp}) are given by
\begin{eqnarray}
      &&  \Big( \frac{R^2 N^{\theta \prime}}{\ell} \Big)^2 
                      + \frac{2(NR')'}{\ell^2} \cdot \frac{2RN}{\ell^2} + \Big( \frac{2NR'}{\ell^2} \Big)^2                           
              =  \frac{\nu^2 + 3}{4 \ell^2} + \frac{9(\nu^2 - 1)^2}{16 \ell^2} \frac{r^2}{R^2}         \nonumber    \\                   
                 && \qquad     - \frac{3(\nu^2 - 1) \sqrt{r_+ r_- (\nu^2 + 3)}}{8 \ell^2} \frac{1}{R^2} ( 2\nu r - \sqrt{r_+ r_- (\nu^2 + 3)} )     \nonumber   \\
      && \qquad + \frac{3(\nu^2 - 1)(\nu^2 + 3)}{16 \ell^2} \frac{1}{R^2} \big(4r^2 - 3(r_+ + r_-) r + 2 r_+ r_- \big)   \,,   \nonumber  \\   
%
      &&  \Big( \frac{R^2 N^{\theta \prime}}{\ell} \Big)' \cdot \frac{2RN}{\ell^2} + \frac{2NR'}{\ell^2} \cdot \frac{2R^2N^{\theta \prime}}{\ell}    \nonumber  \\
      && \quad = - \frac{3(\nu^2 - 1) \sqrt{(\nu^2 + 3)}}{4 \ell^2}   
                   \cdot \frac{1}{R^2} \big( 2\nu r - \sqrt{r_+ r_- (\nu^2 + 3)} \big) \sqrt{(r - r_+)(r - r_-)}  \,,   \nonumber \\
%
      &&   \Big( \frac{R^2 N^{\theta \prime}}{\ell} \Big)^2 + \frac{2RN'}{\ell^2} \frac{2NR'}{\ell^2}              
            =  \frac{\nu^2 + 3}{4 \ell^2} + \frac{9(\nu^2 - 1)^2}{16 \ell^2} \frac{r^2}{R^2}                      \\     
      && \qquad     - \frac{3(\nu^2 - 1) \sqrt{r_+ r_- (\nu^2 + 3)}}{8 \ell^2} \frac{1}{R^2} ( 2\nu r - \sqrt{r_+ r_- (\nu^2 + 3)} )    \nonumber   \\
      && \qquad + \frac{3(\nu^2 - 1)(\nu^2 + 3)}{16 \ell^2} \frac{1}{R^2} \big((r_+ + r_-) r - 2 r_+ r_- \big)   \,,     \nonumber     \\
%
      &&  3 \Big( \frac{R^2 N^{\theta \prime}}{\ell} \Big)^2 - \frac{2(RN')'}{\ell^2} \cdot \frac{2RN}{\ell^2} - \Big( \frac{2RN'}{\ell^2} \Big)^2  
              = - \frac{\nu^2 + 3}{4 \ell^2} + 3 \frac{9(\nu^2 - 1)^2}{16 \ell^2} \frac{r^2}{R^2}        \nonumber  \\               
      && \qquad   - 3 \frac{3(\nu^2 - 1) \sqrt{r_+ r_- (\nu^2 + 3)}}{8 \ell^2} \frac{1}{R^2} ( 2\nu r - \sqrt{r_+ r_- (\nu^2 + 3)} )       \nonumber         \\
      && \qquad + \frac{3(\nu^2 - 1)(\nu^2 + 3)}{16 \ell^2} \cdot \frac{1}{R^2} \big(4r^2 - (r_+ + r_-) r - 2 r_+ r_- \big)    \,.    \nonumber      
\end{eqnarray}
Substituting all the above results into (\ref{cur-warp}) we can rearrange curvature 2-forms for simple forms
\begin{eqnarray} \label{cur-fin}
     && R^0 = \Big( \frac{\nu^2}{\ell^2} + F(r) \Big) e^1 \wedge e^2 + G(r) e^1 \wedge e^0 \,,    \nonumber  \\
     && R^1 = - \frac{\nu^2}{\ell^2} \, e^2 \wedge e^0 \,,            \\
     && R^2 = \Big( - \frac{\nu^2}{\ell^2} + \frac{3(\nu^2 - 1)}{\ell^2} + F(r) \Big) e^0 \wedge e^1 - G(r) e^1 \wedge e^2  \,,    \nonumber
\end{eqnarray}
where
\begin{eqnarray*}
       F(r) &=& \frac{3(\nu^2 - 1)(\nu^2 + 3)}{4 \ell^2} \cdot \frac{1}{R^2} (r - r_+)(r - r_-) \,,     \\
       G(r) &=& - \frac{3(\nu^2 - 1) \sqrt{(\nu^2 + 3)}}{4 \ell^2}        
                       \cdot \frac{1}{R^2} \big( 2\nu r - \sqrt{r_+ r_- (\nu^2 + 3)} \big) \sqrt{(r - r_+)(r - r_-)}    \,.
\end{eqnarray*}
With (\ref{Funcs}) and (\ref{Derivs}) we can calculate the coefficients of the dreibeins in (\ref{aux-warp}), then  the auxiliary fields $h^a$ are simply given by   
\begin{eqnarray} \label{auxs-wap}
         && h^0 = \frac{1}{2\mu} \Big( \frac{\nu^2}{\ell^2} - \frac{3(\nu^2 - 1)}{\ell^2} - 2F \Big) e^0 - \frac{1}{\mu} G e^2 \,,  \nonumber  \\
         && h^1 = \frac{1}{2\mu} \Big( \frac{\nu^2}{\ell^2} - \frac{3(\nu^2 - 1)}{\ell^2} \Big) e^1 \,,              \\
         && h^2 = \frac{1}{2\mu} \Big( \frac{\nu^2}{\ell^2} + \frac{3(\nu^2 - 1)}{\ell^2} + 2F \Big) e^2 + \frac{1}{\mu} G e^0   \,.    \nonumber
\end{eqnarray}  
%

\renewcommand{\theequation}{B.\arabic{equation}}
\setcounter{equation}{0}

\section{Some useful formulae for the charge variations of the warped $AdS_3$ black hole in Topologically Massive Gravity}

To calculate the charge variation for the warped $AdS_3$ black hole we should make the variation of some functions appeared in eq.(\ref{c1-var-wap}). 
The useful variations for (\ref{c1-var-wap}) is as follows
\begin{eqnarray}
     \delta R &=& \frac{(\nu^2 + 3)}{8} \frac{r}{R} (\delta r_+ + \delta r_-)   
                      - \frac{\nu}{4} \frac{r}{R} \frac{\sqrt{r_+ r_- (\nu^2 + 3)}}{r_+ r_-} (r_- \delta r_+ + r_+ \delta r_-)  \,,     \nonumber     \\
%
     \delta(NRR') &=& \frac{\ell \sqrt{(\nu^2 + 3)}}{2} \bigg\{ - \Big( \frac{R}{2r} + \frac{3(\nu^2 - 1)}{8} \frac{r}{R} \Big)    
                             \cdot   \frac{\big[ (r - r_-) \delta r_+ + (r - r_+) \delta r_- \big]}{2\sqrt{(r - r_+)(r - r_-)}}      \nonumber    \\
                      &&  + \Big( \frac{1}{2r} - \frac{3(\nu^2 - 1)}{8} \frac{r}{R^2} \Big) \sqrt{(r - r_+)(r - r_-)}  \delta R  \bigg\}  \,,    \nonumber    \\
%
     \delta(R^3 N^{\theta \prime}) &=& \Big( \frac{R}{2r} + \frac{3(\nu^2 - 1)}{8} \frac{r}{R} \Big)     
                         \cdot    \frac{\sqrt{r_+ r_- (\nu^2 + 3)}}{2 r_+ r_-} (r_- \delta r_+ + r_+ \delta r_-)           \\
                      && + \bigg\{ \frac{\sqrt{r_+ r_- (\nu^2 + 3)}}{2r}    
                            + \frac{3(\nu^2 - 1)}{8} \frac{r}{R^2} (2\nu r - \sqrt{r_+ r_- (\nu^2 + 3)}) \bigg\} \delta R   \,,    \nonumber    \\
%
     \delta(G R) &=& \frac{3(\nu^2 - 1)\sqrt{(\nu^2 + 3)}}{4 \ell^2} \bigg\{ \frac{1}{R^2} (2\nu r - \sqrt{r_+ r_- (\nu^2 + 3)})   
                           \cdot  \sqrt{(r - r_+)(r - r_-)} \delta R  \nonumber  \\
                     && + \frac{\sqrt{(r - r_+)(r - r_-)}}{R} \frac{\sqrt{r_+ r_- (\nu^2 + 3)}}{2 r_+ r_-}  
                           \cdot   ( r_- \delta r_+ + r_+ \delta r_- )      \nonumber     \\
                     && + \frac{2\nu r - \sqrt{r_+ r_- (\nu^2 + 3)}}{2R \sqrt{(r - r_+)(r - r_-)}}    
                           \cdot \big[ (r - r_-) \delta r_+ + (r - r_+) \delta r_- \big] \bigg\}  \,,   \nonumber       \\
%
     \delta(F R) &=& \frac{3(\nu^2 - 1)(\nu^2 + 3)}{4 \ell^2} \bigg\{ - \frac{(r - r_+)(r - r_-)}{R^2} \delta R     
                         - \frac{1}{R} \big[ (r - r_- ) \delta r_+ + (r - r_+) \delta r_-  \big] \bigg\}  \,.      \nonumber 
\end{eqnarray}
In order to get the charge variation (\ref{c1-var-wap}) and (\ref{c2-var-wap}), we consider the variations of frame fields, i.e. dreibein, connections 1-forms
and auxiliary fields. The charge variations are defined by the integral over a certain Cauchy surface with constant time and two boundaries at horizon surface and infinity. So, we only need to consider the variation of the angle $\theta$ part, i.e. $d\theta$ terms. The non-vanishing variations for dreibein (\ref{drei-wap}) and connection 1-forms (\ref{con-wap}) are as follows 
\begin{equation} \label{var-e-w}
      \delta e^2 = \ell \delta R d\theta \,, ~  \delta \omega^0 = \frac{2}{\ell} \delta(NRR') d\theta \,,  ~
      \delta \omega^2 = - \delta(R^3 N^{\theta \prime}) d\theta  \,.
\end{equation}
The variations of auxiliary fields (\ref{auxs-wap}) are given by
\begin{equation}\label{var-aux-wap}
    \delta h^0 = - \frac{\ell}{\mu} \delta(G R) d\theta  \,,   \quad
    \delta h^2 = \bigg\{ \frac{\ell}{2\mu} \Big( \frac{\nu^2}{\ell^2} + \frac{3(\nu^2 - 1)}{\ell^2} \Big) \delta R + \frac{\ell}{\mu} \delta(F R) \bigg\} d\theta   \,. 
\end{equation}

To find the black hole mass variations we should find some interior product of dreibeins, connection 1-forms and auxiliary fields with $\xi = \frac{\partial}{\partial t}$. 
The non-vanishing interior products of dreibeins and connections are given by
\begin{eqnarray}
       &&    i_{\xi} e^0 = N \,,      \quad
                      i_{\xi} e^2 = \ell RN^{\theta} \,,      \nonumber   \\
        &&  i_{\xi} \omega^0 = \frac{R^2 N^{\theta \prime}}{\ell} N + \frac{2NR'}{\ell^2}  \ell R N^{\theta} \,,  \quad    
          i_{\xi} \omega^2 = - \frac{R^2 N^{\theta \prime}}{\ell}  \ell R N^{\theta} + \frac{2RN'}{\ell^2} N \,,   
\end{eqnarray}
and those of auxiliary fields are given by
\begin{eqnarray}
    && i_{\xi} h^0 = \frac{1}{2\mu} \Big( \frac{\nu^2}{\ell^2} - \frac{3(\nu^2 - 1)}{\ell^2} - 2F \Big) N - \frac{1}{\mu} G \ell R N^{\theta}  \,,     \nonumber  \\
    && i_{\xi} h^2 = \frac{1}{2\mu} \Big( \frac{\nu^2}{\ell^2} + \frac{3(\nu^2 - 1)}{\ell^2} +2F \Big) \ell RN^{\theta} + \frac{1}{\mu} G N \,.     
\end{eqnarray}
To find the black hole angular momentum variations with $\xi = \frac{\partial}{\partial \theta}$ the non-vanishing interior products are given by
\begin{equation}
       i_{\xi} e^2 = \ell R \,, \quad  i_{\xi} \omega^0 = \frac{2NR'}{\ell^2} \ell R \,, \quad
       i_{\xi} \omega^2 = - \frac{R^2 N^{\theta \prime}}{\ell} \ell R \,, 
\end{equation}
and 
\begin{equation}
     i_{\xi} h^0 = - \frac{\ell}{\mu} G R  \,,   \quad
     i_{\xi} h^2 = \frac{\ell}{2\mu} \Big( \frac{\nu^2}{\ell^2} + \frac{3(\nu^2 - 1)}{\ell^2} +2F \Big) R  \,.
\end{equation}
To compute the angular momentum (\ref{Ang-C1}) we need to use the following some functions
\begin{eqnarray}
    R^4 N^{\theta \prime} &=& - \frac{3\nu (\nu^2 - 1)}{4} r^2     
              + R \Big( \frac{R}{2r} + \frac{3(\nu^2 - 1)}{8} \frac{r}{R} \Big) \sqrt{r_+ r_- (\nu^2 + 3)} \,,     \nonumber   \\
    (NRR')^2 &=& \frac{\ell^2 (\nu^2 + 3)}{4} (r - r_+)(r - r_-)        
                \cdot \Big( \frac{R}{2r} + \frac{3(\nu^2 - 1)}{8} \frac{r}{R} \Big)^2   \,,     \\
    (R^3 N^{\theta \prime})^2 &=& \bigg\{ - \frac{3\nu (\nu^2 - 1)}{4} \frac{r^2}{R}      
              + \Big( \frac{R}{2r} + \frac{3(\nu^2 - 1)}{8} \frac{r}{R} \Big) \sqrt{r_+ r_- (\nu^2 + 3)} \bigg\}^2   \,,   \nonumber  \\
    R^2 F &=& \frac{3(\nu^2 - 1)(\nu^2 + 3)}{4\ell^2} (r - r_+)(r - r_-) \,.       \nonumber   
\end{eqnarray}
Owing to the above formulae the charge variation for the angular momentum (\ref{Ang-C1}) can be expressed by the following form 
\begin{eqnarray}
     \chi_{\xi} \Big[ \frac{\partial}{\partial \theta} \Big] 
            &=& \bigg\{ \sigma \ell \bigg( - \frac{3\nu (\nu^2 - 1)}{4} r^2      
                    + R \Big( \frac{R}{2r} + \frac{3(\nu^2 - 1)}{8} \frac{r}{R} \Big) \sqrt{r_+ r_- (\nu^2 + 3)} \bigg)    \nonumber  \\
              && - \frac{2}{\mu \ell^2} \frac{\ell^2 (\nu^2 + 3)}{4} (r - r_+)(r - r_-) \Big( \frac{R}{2r} + \frac{3(\nu^2 - 1)}{8} \frac{r}{R} \Big)^2   \nonumber     \\
              && + \frac{1}{2\mu} \bigg( - \frac{3\nu (\nu^2 - 1)}{4} \frac{r^2}{R}     
                    + \Big( \frac{R}{2r} + \frac{3(\nu^2 - 1)}{8} \frac{r}{R} \Big) \sqrt{r_+ r_- (\nu^2 + 3)} \bigg)^2   \nonumber    \\
              && + \frac{1}{2\mu} (\nu^2 + 3(\nu^2 - 1)) R^2      
                    + \frac{\ell^2}{\mu} \frac{3(\nu^2 - 1)(\nu^2 + 3)}{4 \ell^2} (r - r_+)(r - r_-) \bigg\} d\theta  \,.  \label{Ang-Q}
\end{eqnarray}
As $r$ goes to the infinity the previous result can be expanded as a series expansion of $r$. 
The useful asymptotic expansions including $R(r)$ described by (\ref{wap-func}) are given by
\begin{eqnarray*}
   && \frac{r}{R} \simeq \frac{2}{\sqrt{3(\nu^2 - 1)}} \Big( 1 - \frac{1}{2r} {\cal C} + \frac{3}{8 r^2} {\cal C}^2 + \cdots \Big)  \,,   \\
   && \frac{R}{r} \simeq \frac{\sqrt{3(\nu^2 - 1)}}{2} \Big( 1 + \frac{1}{2r} {\cal C} - \frac{1}{8r^2} {\cal C}^2 + \cdots \Big)   \,,   \\
   && \Big( \frac{R}{2r} + \frac{3(\nu^2 - 1)}{8} \frac{r}{R} \Big) \simeq 
             \frac{\sqrt{3(\nu^2 - 1)}}{2} \Big( 1 + \frac{1}{8r^2} {\cal C}^2 \cdots \Big)  \,,
\end{eqnarray*}
where
\begin{equation*}
     {\cal C} = \frac{\nu^2 + 3}{3(\nu^2 - 1)} (r_+ + r_-) - \frac{4\nu}{3(\nu^2 - 1)} \sqrt{r_+ r_- (\nu^2 + 3)} \,.
\end{equation*}
Using the above approximations we can re-express the charge (\ref{Ang-Q}) with $\sigma = 1$ and $\frac{1}{\mu} = \frac{\ell}{3\nu}$. 
So, rearranging the charge with respect to $r$ orders  becomes 
\begin{eqnarray}
   \chi_{\xi} \Big[ \frac{\partial}{\partial \theta} \Big]
                &=& \bigg\{ \ell \frac{3(\nu^2 - 1)}{8} \sqrt{r_+ r_- (\nu^2 + 3)} {\cal C}      
                        - \frac{\ell}{3\nu} \frac{3(\nu^2 - 1)(\nu^2 + 3)}{8} \Big( \frac{1}{4} {\cal C}^2 + r_+ r_- \Big)   \nonumber    \\
                && + \frac{\ell}{3\nu} \frac{3\nu^2 (\nu^2 - 1)}{8} \Big( {\cal C}^2 + \frac{1}{\nu} \sqrt{r_+ r_- (\nu^2 + 3)} {\cal C}   
                      + \frac{r_+ r_- (\nu^2 + 3)}{\nu^2} \Big)              \nonumber  \\
                && + \frac{\ell}{3\nu} \frac{3(\nu^2 - 1)(\nu^2 + 3)}{4} r_+ r_- \bigg\} d\theta    
                      + {\cal O} \Big( \frac{1}{r} \Big) d\theta   \,,            \label{Chi-Ang-wap}
\end{eqnarray}
where the coefficients of $r^2$ and $r$ are vanished. Therefore it gives the result (\ref{Chi-Ang}).

\newpage


\begin{thebibliography}{99}



\bibitem{Deser:1983tn} 
  S.~Deser, R.~Jackiw and G.~'t Hooft,
  Annals Phys.\  {\bf 152}, 220 (1984).
  
  
  
\bibitem{Banados:1992wn} 
  M.~Banados, C.~Teitelboim and J.~Zanelli,
  Phys.\ Rev.\ Lett.\  {\bf 69}, 1849 (1992)
  [hep-th/9204099].  
  


\bibitem{Banados:1992gq} 
  M.~Banados, M.~Henneaux, C.~Teitelboim and J.~Zanelli,
  Phys.\ Rev.\ D {\bf 48}, 1506 (1993)
  Erratum: [Phys.\ Rev.\ D {\bf 88}, 069902 (2013)]
  [gr-qc/9302012].



\bibitem{Deser:1981wh} 
  S.~Deser, R.~Jackiw and S.~Templeton,
  Annals Phys.\  {\bf 140}, 372 (1982)
  [Annals Phys.\  {\bf 281}, 409 (2000)]
  Erratum: [Annals Phys.\  {\bf 185}, 406 (1988)].
  
  
  
\bibitem{Deser:1981wh-1}
  S.~Deser, R.~Jackiw and S.~Templeton,
  Annals Phys.\  {\bf 281}, 409 (2000)




\bibitem{Deser:1982}
  R.~Jackiw, S.~Templeton and S.~Deser,
  Phys.\ Rev.\ Lett. {\bf48} (1982) 975.


\bibitem{Bouchareb:2007yx} 
  A.~Bouchareb and G.~Clement,
  Class.\ Quant.\ Grav.\  {\bf 24}, 5581 (2007)
  [arXiv:0706.0263 [gr-qc]].



\bibitem{Hotta:2008yq} 
  K.~Hotta, Y.~Hyakutake, T.~Kubota and H.~Tanida,
  JHEP {\bf 0807}, 066 (2008)
  [arXiv:0805.2005 [hep-th]].



\bibitem{Skenderis:2009nt} 
  K.~Skenderis, M.~Taylor and B.~C.~van Rees,
  JHEP {\bf 0909}, 045 (2009)
  [arXiv:0906.4926 [hep-th]].



\bibitem{Henningson:1998gx} 
  M.~Henningson and K.~Skenderis,
  JHEP {\bf 9807}, 023 (1998)
  [hep-th/9806087].




\bibitem{Balasubramanian:1999re} 
  V.~Balasubramanian and P.~Kraus,
  Commun.\ Math.\ Phys.\  {\bf 208}, 413 (1999)
  [hep-th/9902121].



\bibitem{deHaro:2000vlm} 
  S.~de Haro, S.~N.~Solodukhin and K.~Skenderis,
  Commun.\ Math.\ Phys.\  {\bf 217}, 595 (2001)
   [hep-th/0002230].





\bibitem{Bergshoeff:2009hq} 
  E.~A.~Bergshoeff, O.~Hohm and P.~K.~Townsend,
  Phys.\ Rev.\ Lett.\  {\bf 102}, 201301 (2009)
   [arXiv:0901.1766 [hep-th]].




\bibitem{Bergshoeff:2009aq} 
  E.~A.~Bergshoeff, O.~Hohm and P.~K.~Townsend,
  Phys.\ Rev.\ D {\bf 79}, 124042 (2009)
  [arXiv:0905.1259 [hep-th]].




\bibitem{Clement:2009gq} 
  G.~Clement,
  Class.\ Quant.\ Grav.\  {\bf 26}, 105015 (2009)
   [arXiv:0902.4634 [hep-th]].



\bibitem{Oliva:2009ip} 
  J.~Oliva, D.~Tempo and R.~Troncoso,
  JHEP {\bf 0907}, 011 (2009)
  [arXiv:0905.1545 [hep-th]].



\bibitem{Giribet:2009qz} 
  G.~Giribet, J.~Oliva, D.~Tempo and R.~Troncoso,
  Phys.\ Rev.\ D {\bf 80}, 124046 (2009)
  [arXiv:0909.2564 [hep-th]].



\bibitem{Ghodsi:2010gk} 
  A.~Ghodsi and M.~Moghadassi,
  Phys.\ Lett.\ B {\bf 695}, 359 (2011)
  [arXiv:1007.4323 [hep-th]].












\bibitem{Maldacena:1997re-1} 
  J.~M.~Maldacena,
  Int.\ J.\ Theor.\ Phys.\  {\bf 38}, 1113 (1999).



\bibitem{Maldacena:1997re} 
  J.~M.~Maldacena,
  Adv.\ Theor.\ Math.\ Phys.\  {\bf 2}, 231 (1998)
  [hep-th/9711200].




\bibitem{Witten:1998qj} 
  E.~Witten,
  Adv.\ Theor.\ Math.\ Phys.\  {\bf 2}, 253 (1998)
  [hep-th/9802150].



\bibitem{Gubser:1998bc} 
  S.~S.~Gubser, I.~R.~Klebanov and A.~M.~Polyakov,
  Phys.\ Lett.\ B {\bf 428}, 105 (1998)
  [hep-th/9802109].



\bibitem{Sinha:2010ai} 
  A.~Sinha,
  JHEP {\bf 1006}, 061 (2010)
  [arXiv:1003.0683 [hep-th]].



\bibitem{Gullu:2010pc} 
  I.~Gullu, T.~C.~Sisman and B.~Tekin,
  Class.\ Quant.\ Grav.\  {\bf 27}, 162001 (2010)
   [arXiv:1003.3935 [hep-th]].



\bibitem{Myers:2010tj} 
  R.~C.~Myers and A.~Sinha,
  JHEP {\bf 1101}, 125 (2011)
   [arXiv:1011.5819 [hep-th]].



\bibitem{Boulware:1973my} 
  D.~G.~Boulware and S.~Deser,
  Phys.\ Rev.\ D {\bf 6}, 3368 (1972).


\bibitem{Bergshoeff:2013xma} 
  E.~A.~Bergshoeff, S.~de Haan, O.~Hohm, W.~Merbis and P.~K.~Townsend,
  Phys.\ Rev.\ Lett.\  {\bf 111}, no. 11, 111102 (2013)
  Erratum: [Phys.\ Rev.\ Lett.\  {\bf 111}, no. 25, 259902 (2013)]
   [arXiv:1307.2774].


\bibitem{Bergshoeff:2014bia} 
  E.~A.~Bergshoeff, O.~Hohm, W.~Merbis, A.~J.~Routh and P.~K.~Townsend,
  Lect.\ Notes Phys.\  {\bf 892}, 181 (2015)
  [arXiv:1402.1688 [hep-th]].



\bibitem{Bergshoeff:2014pca} 
  E.~Bergshoeff, O.~Hohm, W.~Merbis, A.~J.~Routh and P.~K.~Townsend,
  Class.\ Quant.\ Grav.\  {\bf 31}, 145008 (2014)
  [arXiv:1404.2867 [hep-th]].



\bibitem{Afshar:2014ffa} 
  H.~R.~Afshar, E.~A.~Bergshoeff and W.~Merbis,
  JHEP {\bf 1408}, 115 (2014)
  [arXiv:1405.6213 [hep-th]].



\bibitem{Arvanitakis:2014yja} 
  A.~S.~Arvanitakis, A.~J.~Routh and P.~K.~Townsend,
  Class.\ Quant.\ Grav.\  {\bf 31}, no. 23, 235012 (2014)
    [arXiv:1407.1264 [hep-th]].



\bibitem{Arvanitakis:2014xna} 
  A.~S.~Arvanitakis and P.~K.~Townsend,
  Class.\ Quant.\ Grav.\  {\bf 32}, no. 8, 085003 (2015)
  [arXiv:1411.1970 [hep-th]].



\bibitem{Merbis:2014vja} 
  W.~Merbis, PhD. thesis university of Groningen,
  arXiv:1411.6888 [hep-th].



\bibitem{Arnowitt:1962hi} 
  R.~L.~Arnowitt, S.~Deser and C.~W.~Misner,
  Gen.\ Rel.\ Grav.\  {\bf 40}, 1997 (2008)
  [gr-qc/0405109].



\bibitem{Brown:1992br} 
  J.~D.~Brown and J.~W.~York, Jr.,
  Phys.\ Rev.\ D {\bf 47}, 1407 (1993)
   [gr-qc/9209012].



\bibitem{Abbott:1981ff} 
  L.~F.~Abbott and S.~Deser,
  Nucl.\ Phys.\ B {\bf 195}, 76 (1982).



\bibitem{Deser:2002rt} 
  S.~Deser and B.~Tekin,
  Phys.\ Rev.\ Lett.\  {\bf 89}, 101101 (2002)
  [hep-th/0205318].




\bibitem{Deser:2002jk} 
  S.~Deser and B.~Tekin,
  Phys.\ Rev.\ D {\bf 67}, 084009 (2003)
  [hep-th/0212292].




\bibitem{Wald:1993nt} 
  R.~M.~Wald,
  Phys.\ Rev.\ D {\bf 48}, 3427 (1993)
  [gr-qc/9307038].





\bibitem{Jacobson:1993vj} 
  T.~Jacobson, G.~Kang and R.~C.~Myers,
  Phys.\ Rev.\ D {\bf 49}, 6587 (1994)
  [gr-qc/9312023].





\bibitem{Iyer:1994ys} 
  V.~Iyer and R.~M.~Wald,
  Phys.\ Rev.\ D {\bf 50}, 846 (1994)
  [gr-qc/9403028].




\bibitem{Wald:1999wa} 
  R.~M.~Wald and A.~Zoupas,
  Phys.\ Rev.\ D {\bf 61}, 084027 (2000)
  [gr-qc/9911095].





\bibitem{Anninos:2008fx} 
  D.~Anninos, W.~Li, M.~Padi, W.~Song and A.~Strominger,
  JHEP {\bf 0903}, 130 (2009)
   [arXiv:0807.3040 [hep-th]].





\bibitem{Kim:2013zha} 
  W.~Kim, S.~Kulkarni and S.~H.~Yi,
  Phys.\ Rev.\ Lett.\  {\bf 111}, no. 8, 081101 (2013)
  Erratum: [Phys.\ Rev.\ Lett.\  {\bf 112}, no. 7, 079902 (2014)]
    [arXiv:1306.2138 [hep-th]].




\bibitem{Kim:2013cor} 
  W.~Kim, S.~Kulkarni and S.~H.~Yi,
  Phys.\ Rev.\ D {\bf 88}, no. 12, 124004 (2013)
    [arXiv:1310.1739 [hep-th]].






\bibitem{Miskovic:2009kr} 
  O.~Miskovic and R.~Olea,
  JHEP {\bf 0912}, 046 (2009)
    [arXiv:0909.2275 [hep-th]].



\bibitem{Giribet:2010ed} 
  G.~Giribet and M.~Leston,
  JHEP {\bf 1009}, 070 (2010)
  [arXiv:1006.3349 [hep-th]].



\bibitem{Hohm:2010jc} 
  O.~Hohm and E.~Tonni,
  JHEP {\bf 1004}, 093 (2010)
  [arXiv:1001.3598 [hep-th]].


\bibitem{Nam:2010dd} 
  S.~Nam, J.~D.~Park and S.~H.~Yi,
  JHEP {\bf 1007}, 058 (2010)
   [arXiv:1005.1619 [hep-th]].



\bibitem{Nam:2010ub} 
  S.~Nam, J.~D.~Park and S.~H.~Yi,
  Phys.\ Rev.\ D {\bf 82}, 124049 (2010)
  [arXiv:1009.1962 [hep-th]].



\bibitem{Deser:2003vh} 
  S.~Deser and B.~Tekin,
  Class.\ Quant.\ Grav.\  {\bf 20}, L259 (2003)
    [gr-qc/0307073].




\bibitem{Clement:2003sr} 
  G.~Clement,
  Phys.\ Rev.\ D {\bf 68}, 024032 (2003)
  [gr-qc/0301129].





\bibitem{Moussa:2003fc} 
  K.~A.~Moussa, G.~Clement and C.~Leygnac,
  Class.\ Quant.\ Grav.\  {\bf 20}, L277 (2003)
    [gr-qc/0303042].

 
 


\bibitem{Alishahiha:2010bw} 
  M.~Alishahiha and A.~Naseh,
  Phys.\ Rev.\ D {\bf 82}, 104043 (2010)
  [arXiv:1005.1544 [hep-th]].





\bibitem{Kwon:2011jz} 
  Y.~Kwon, S.~Nam, J.~D.~Park and S.~H.~Yi,
  JHEP {\bf 1111}, 029 (2011)
  [arXiv:1106.4609 [hep-th]].





\bibitem{Tekin:2014jna} 
  B.~Tekin,
  Phys.\ Rev.\ D {\bf 90}, no. 8, 081701 (2014)
  [arXiv:1409.5358 [hep-th]].




\bibitem{Setare:2015pva} 
  M.~R.~Setare and H.~Adami,
  Phys.\ Rev.\ D {\bf 91}, no. 10, 104039 (2015)
  [arXiv:1501.00920 [hep-th]].




\bibitem{Tachikawa:2006sz} 
  Y.~Tachikawa,
  Class.\ Quant.\ Grav.\  {\bf 24}, 737 (2007)
   [hep-th/0611141].
  
  

\bibitem{Setare:2014zea} 
  M.~R.~Setare,
  Nucl.\ Phys.\ B {\bf 898}, 259 (2015)
   [arXiv:1412.2151 [hep-th]].



\bibitem{Setare:2015vea} 
  M.~R.~Setare and H.~Adami,
  Phys.\ Lett.\ B {\bf 744}, 280 (2015)
  [arXiv:1504.01660 [gr-qc]].

 


\bibitem{Alishahiha:2014dma} 
  M.~Alishahiha, M.~M.~Qaemmaqami, A.~Naseh and A.~Shirzad,
  JHEP {\bf 1412}, 033 (2014)
  [arXiv:1409.6146 [hep-th]].
  
  
  
  
  
\bibitem{Alishahiha:2015whv} 
  M.~Alishahiha, M.~M.~Qaemmaqami, A.~Naseh and A.~Shirzad,
  JHEP {\bf 1601}, 106 (2016)
  [arXiv:1511.06194 [hep-th]].  






\bibitem{Jacobson:2015uqa} 
  T.~Jacobson and A.~Mohd,
  Phys.\ Rev.\ D {\bf 92}, 124010 (2015)
  [arXiv:1507.01054 [gr-qc]].
  
 
 
 
\bibitem{Setare:2015nla} 
  M.~R.~Setare and H.~Adami,
  Nucl.\ Phys.\ B {\bf 902}, 115 (2016)
  [arXiv:1509.05972 [hep-th]].
  
  
  
  
\bibitem{Setare:2015cvv} 
  H.~Adami and M.~R.~Setare,
  Eur.\ Phys.\ J.\ C {\bf 76}, no. 4, 187 (2016)
  [arXiv:1511.00527 [gr-qc]].
  
  
  
  

\bibitem{Setare:2015gss} 
  M.~R.~Setare and H.~Adami,
  Nucl.\ Phys.\ B {\bf 909}, 345 (2016)
  [arXiv:1511.01070 [hep-th]].
  
  
  
  
  
\bibitem{Setare:2016hmn} 
  M.~R.~Setare and H.~Adami,
  Nucl.\ Phys.\ B {\bf 909}, 297 (2016)
  [arXiv:1601.00171 [hep-th]].  
    
 
 
 
 
 
\bibitem{Cardy:1986ie} 
  J.~L.~Cardy,
  Nucl.\ Phys.\ B {\bf 270}, 186 (1986).
 
 
 
 
 
\bibitem{Carlip:1998qw} 
  S.~Carlip,
  Class.\ Quant.\ Grav.\  {\bf 15}, 3609 (1998)
  [hep-th/9806026].
 
 
 
 
 
\bibitem{Hohm:2012vh} 
  O.~Hohm, A.~Routh, P.~K.~Townsend and B.~Zhang,
  Phys.\ Rev.\ D {\bf 86}, 084035 (2012)
  [arXiv:1208.0038 [hep-th]].
 
 
 
 
 
 
 
\bibitem{Blagojevic:2010ir} 
  M.~Blagojevic and B.~Cvetkovic,
  JHEP {\bf 1101}, 082 (2011)
  [arXiv:1010.2596 [gr-qc]].





\bibitem{Carlip:2008qh} 
  S.~Carlip,
  JHEP {\bf 0810}, 078 (2008)
  [arXiv:0807.4152 [hep-th]].






\bibitem{Brown:1986nw} 
  J.~D.~Brown and M.~Henneaux,
  Commun.\ Math.\ Phys.\  {\bf 104}, 207 (1986).




\bibitem{Adami:2017phg} 
  H.~Adami, M.~R.~Setare, T.~C.~Sisman and B.~Tekin, A review paper,
  [arXiv:1710.07252 [hep-th]].


 
     
 
 
 
 
 
 
 







\end{thebibliography}
\end{document}